\documentclass[12pt]{article}

\usepackage{amssymb}
\usepackage{amsmath}
\usepackage{graphicx, color, psfrag, epsfig}
\usepackage{enumitem}
\usepackage{natbib}
\usepackage{url} 
\usepackage{verbatim} 
\usepackage{caption}

\global\hyphenpenalty=0

\def\@fnsymbol#1{\ensuremath{\ifcase#1\or *\or \dagger\or \ddagger\or
   \mathsection\or \mathparagraph\or \|\or **\or \dagger\dagger
   \or \ddagger\ddagger \else\@ctrerr\fi}}
   
\usepackage{changepage}

\begin{document}

\newcommand{\dblprime}{^{\prime\prime}}

\captionsetup[figure]{labelsep=space}
 
\renewcommand{\thefootnote}{\fnsymbol{footnote}}
 
\changepage{+1.5in}{}{}{}{}{-1.2in}{}{}{}
 
\begin{titlepage}
    \begin{center}
        \vspace*{1cm}         
        \LARGE
        \textbf{Economic DAO Governance:\\ A Contestable Control Approach} 
                   
        \vspace{0.5cm}
        \large  
        \textbf{Jeff Strnad\footnote{Charles A. Beardsley Professor of Law, Stanford University. Correspondence email: jeffstrnad5@gmail.com. I am grateful for valuable comments from Bobby Bartlett, Dan Boneh, Albert Choi, Silke Elrifai, Mike Guttentag, Jake Hartnell, Henry Hu, Takuma Iwasaki, Mark Lemley, Sven Riva, Roberta Romano, Martin Schmidt, Mike Simkovic, Connor Spelliscy, and Jay Yu, and from participants at the DAO @ LexTech Institute 2024 Conference, the 2024 Stanford Blockchain Governance Summit, the Stanford Law School Faculty Workshop, the TLDR Conference 2024, the 2024 American Law and Economics Association Annual Meeting, Edge Esmeralda 2024, UBRI Connect 2024, and the Hackers Congress Paraleln\'i Polis 2024. I have also benefited greatly from conversations with Andy Hall.}}

\thispagestyle{empty}   

\vspace{8mm}

\begin{abstract}
In this article, we propose a new form of DAO governance that uses a
sequential auction mechanism to overcome entrenched control issues that have
emerged for DAOs by creating a regime of temporary contest\-able control. The
mechanism avoids potential public choice problems inherent in voting
approaches but at the same time provides a vehicle that can enhance and
secure value that inheres to DAO voting and other DAO non-market governance
procedures. It is robust to empty voting and is code feasible. It not only facilitates
the ability of DAOs to meet their normative and operational goals in the face
of diverse regulatory approaches, but also strengthens the case for creating a less burdensome but at least equally effective regulatory regime for DAOs that employ the mechanism. 
Designed to shift control to the party with
the most promising business plan, at the same time it deters value destruction by control parties, maximizes social surplus, and distributes that surplus in
a way that tends to promote invest\-ment by
other parties both at start up and on an on-going basis.\\

\vspace{4mm}
\end{abstract}

\vspace{8mm} 

\normalsize


November 25, 2024

\normalsize

\vspace{4mm}

\copyright Jeff Strnad

\vspace{4mm}

\end{center}

\newpage 

\changepage{-1.5in}{}{}{}{}{+1.2in}{}{}{}


\tableofcontents \thispagestyle{empty}

\end{titlepage}     

\renewcommand*{\thefootnote}{\arabic{footnote}}
\setcounter{footnote}{0}

\changepage{-1.5in}{}{}{}{}{+1.2in}{}{}{}

\section{Introduction}\label{s1}

Decentralized Autonomous Organizations (``DAOs") are a recent innovation,
dating from the April 2016 launch of ``The
DAO."\footnote{\citet{Vigna_2016}.} DAOs operate largely through the
execution of code and have no centralized management. When participant
decisions are required, DAOs typically utilize voting by token holders,
roughly analogous to shareholder voting in a corporate setting. In most
cases, the tokens are publicly traded.

Governance issues for DAOs have received considerable attention of late. Lots
of experimentation with governance is taking place as well as a large volume
of commentary.\footnote{The Decentralization Research Center (formerly, the
DAO Research Collective) website includes a significant and representative
collection of commentary and descriptions of experiments under the headings
``Governance" and ``Decentralization." \citet{DRC_2023}.} Most of the
experimentation has centered around different voting mechanisms, including
quorum-based token voting, a ``direct democracy'' approach in which token
holders vote on proposals subject to a quorum requirement, and approaches in
which delegation or other forms of representation are
possible.\footnote{\citet{Nigam_2023}(section on ``Voting systems").}

Token voting potentially clashes with the goal of decentralization because of
the danger of two types of entrenchment that threaten to create the
equivalent of centralized management. First, \emph{explicit control} is
attainable by accumulating a sufficiently large token position. Second, it is
often the case that a chronic lack of voter participation puts \emph{implicit
control} in the hands of a small set of active token holders who regularly
engage in governance voting.\footnote{See, e.g., \citet{Feichtinger_2023},
\citet{Sun_2022}, and \citet{Liu_2023}.} Chronic lack of participation is
consistent with rationality in many instances. The passive token holders may
be portfolio investors or may have a small enough holding that the costs of
being an informed voter greatly exceed the potential benefits in the form of
higher token values or otherwise, especially in light of the low probability of being the decisive voter with respect to each proposal. More generally, the benefits of any effort
expended by a token holder to become an informed voter will accrue mostly to
other token holders who are in effect free riders.\footnote{\citet[pp.
239-240]{Khanna_2022} describes this collective action problem in the
corporate setting with reference to some of the literature. \citet[p.
26]{Reyes_2017} appear to be the first to note the presence and importance of
the same problem in the DAO setting.}

In order to address the danger of entrenchment as well as several other major
potential problems for DAOs, we propose a different approach to DAO
governance centered on a sequential auction mechanism. The fundamental
building block of the mechanism is a \emph{basic auction} that grants the
auction winner \emph{temporary, contestable control} of the DAO. Bids in the
basic auction consist of a token target price, a surplus claim, and
revelation of any toehold position held by the bidder at the time of the bid.
Under the mechanism, the target token price part of the bid is effectively a
\emph{value claim}, $S$, by the bidder that with control the bidder can
implement a business plan that increases the token value from the prevailing
market price, $P_0$, at the time of the bid to $S$. The \emph{surplus claim} part of
the bid is a claim by the bidder to a portion of the \emph{total added token value},
$(S-P_0)q$, generated if the token target price is attained, where $q$ is the
total number of tokens outstanding. The winning bid is the one that leaves
the largest amount of added token value to the other token holders, that is, the bid that maximizes total added token value minus the sum of the bidder's surplus claim and the expected gain on the bidder's toehold position if the value claim, $S$, is realized.

In section \ref{mechanism} and Appendix A.1, we show that the dominant
strategy in the basic auction is for the bidder to choose a business plan
that produces the largest possible \emph{social surplus}, $(V-P_0)q - C$, the total added token value minus the bidder's cost, and
to truthfully reveal the token value, $V$, that the bidder envisions as
attainable under that business plan. As a result, with some minor exceptions,
the mechanism will choose the socially best project, the one that maximizes
$(V-P_0)q-C$. In addition, because the winning bid is the one that allocates the largest amount of social surplus to the existing token holders, the mechanism will tend to maximize the amount of such surplus realized by token holders from future innovations, which in turn will tend to maximize the market value of the DAO both at start up and on an on-going basis.

The basic auction moves the DAO from a \emph{default governance state} that
typically is a voting regime into a \emph{control period} in which the
winning bidder controls the DAO independent of holding the majority or
supermajority of tokens required to secure control under the voting regime.
This feature allows the other token holders to enjoy a majority or even an
overwhelming majority of the social surplus generated by the winning
bidder's project through their token ownership. Token ownership and control
are separated during the control period.

The sequential aspect of the mechanism ensures that control arising from a
basic auction is temporary. There are several terminating events. If the
control party succeeds in reaching the token target price on a sustained basis, control ends and the DAO reverts to its default governance state. The control period ends if there is a supervening auction and the control party fails to win that auction. Control is contestable because the mechanism allows a supervening auction to take place at any time, triggered by any party willing to make an initial bid. Finally, the control period is limited in time and ends when the limit is reached even if the token target price was not realized and there was no supervening auction.

Section \ref{mechanism} details a set of features that creates appropriate
bidding and project execution incentives for control parties. Control parties
that fail to achieve the token target price forfeit a portion or all of a
substantial \emph{value deposit} to the other token holders, guaranteeing
that those token holders receive the full benefit of the winning bidder's
claimed future performance whether or not execution is successful.
\emph{Value destruction} by control parties is deterred by potential loss of
both the value deposit and an additional \emph{surety deposit}. Value
destruction is a serious concern because control may be separated from
ownership to an extreme degree. A party with control but very few tokens can
take a large financial short position and then tank the DAO. The threat of
losing the value deposit also creates appropriate post-auction incentives to execute the business plan successfully.

The mechanism just described defeats potential entrenchment in the form of
explicit or implicit control. Holding a majority or supermajority of tokens
no longer secures control of the DAO. The majority holder may lose control to
a party with little or no token stake by being outbid in a basic auction, and
a basic auction may be initiated at any time by a party willing to submit an
initial bid. Similarly, implicit control is insecure. A single challenging
party can initiate an auction contest for control even in the presence of
total passivity on the part of the vast majority of token holders, the
passivity that enabled a small group to control the DAO implicitly.

The social value properties of the mechanism emerge definitively only if the
token value of the DAO represents the intrinsic value of the DAO itself. As a
result, the proposed approach is limited to \emph{economic DAOs}, defined as
DAOs with publicly traded governance tokens for which the token value
reflects the inherent value of the enterprise.

DAOs vary greatly in purpose and approach. Some DAOs are very similar to
commercial businesses providing services such as a trading exchange. These
\emph{commercial DAOs} are the most obvious example of economic DAOs. They
provide goods and services with the profits accruing to token holders. The
market capitalization of a commercial DAO represents its value in terms of
the risk-adjusted present value of future returns. But economic DAOs are a
much broader category. All that is necessary is that the market
capitalization represents the value of the enterprise to the token holders. A
charitable or investment DAO, for example, might gather funds to donate or
invest. Such DAOs might have particular objectives, such as promoting
environmental or climate change goals. Greater effectiveness at what they do
in the view of all \emph{potential} participants, including the specification
of the objectives themselves, translates into a higher demand for the tokens
along with a higher market capitalization.\footnote{Exit and entry through
the token trading market is easy and nearly costless for publicly-traded
DAOs. The associated community is open, not limited to current token holders.
As a result, designing a governance system for a publicly traded DAO is very
different from designing such a system for communities such as nation states
or social clubs that are relatively closed because moving in and out of the
community voluntarily is much more costly or even impossible. A range of
non-commercial DAOs can be analogized to cities and towns offering different
mixtures of local public goods in exchange for a different packages of taxes
to residents and potential residents in a hypothetical world characterized by
negligible moving costs.}

An important ideal for a DAO governance mechanism is for most, and ideally
all, aspects to be \emph{code feasible}. Code feasible means implementable
using available blockchain technology without recourse to external
institutions. Code feasibility is a key aspect of decentralization, close to
a necessary condition. By decentralization we mean the ability for the token
project to operate in the absence of trusted parties.

The desideratum of code feasibility distinguishes the case of DAOs from
conventional corporate or public governance structures. Those structures
depend primarily on human management rather than code implementations that
are automatic and, at present, are also more constrained by various legal
requirements and regulations. The mechanism developed here is potentially
applicable to the governance of some conventional institutions, in
particular, public corporations. We leave full consideration of such
applications to future work because they involve their own considerable and
distinct complexities as well as a different, more stringent set of legal
constraints. On the other hand, theoretical work and experience with
governance of conventional institutions is pertinent, and we draw on both in
what follows.

The sequential auction mechanism proposed here eliminates the danger of
entrenchment inherent in token-voting schemes. It also has some very positive
social value aspects: It tends to promote choice and implementation of the
best set of business plans while at the same time securing the highest
possible initial and on-going investment value for the DAO by allocating as
much social surplus as possible to existing token holders. Going further, it is
important to consider how the mechanism relates to voting approaches,
decentralization, and the associated web3 ideals that emphasize the role that
DAOs can play in creating new kinds of democratic communities. Section
\ref{voting} addresses the interaction of the mechanism with voting
approaches, which is possible without first going through a more technical
description of the mechanism. That section has an introductory aspect because
the interaction with voting is one motivation for the sequential aspects of
the mechanism. Section \ref{roots} provides a conceptual overview of the
mechanism with reference to the pre-existing corporate governance literature.
Section \ref{mechanism} together with two Appendices present the full,
technical version of the mechanism, including some evaluative aspects that
arise naturally as part of the presentation. Section \ref{more} contains
further evaluation, including a final subsection discussing decentralization
and closely related regulatory considerations. Section \ref{concluding}
concludes with an assessment of the mechanism in light of web3 ideals.

\section{Interaction of the Mechanism with Voting}\label{voting}

The mechanism enables not only a sequence of auctions for control but also
possible intervening periods in which governance reverts to a default
governance state, typically a voting method of some kind. This approach
allows the mechanism to achieve both operational goals and procedural goals
in any combination or sequence. As discussed previously, at the operational
level, the basic auctions comprising the sequence allow identification of the
best business plans for the DAO combined with a tendency to implement them in
a way that shifts as much social surplus as possible to existing token holders from
the control parties who undertake the implementation. When no party is
willing to bid for control, the mechanism restores the default
governance state.

Aside from operational efficiency, DAO participants may value particular
procedural approaches that embody \emph{process values} in the form of
certain community, ``democratic,'' or participatory norms. Potential control
periods allow for a reset of the default governance state when it departs
from the desired process values or, more ambitiously, revision of the default
governance state itself. In terms of voting methods, the mechanism is
available to preserve process values both by providing a guardrail that
corrects voting method failures in an interim fashion before reinstating the
method and by facilitating comprehensive reform of the voting method itself
if desired by current and potential DAO participants.

Three subsections follow. The first two describe how the mechanism can
correct possible voting method failures that may have negative operational or
procedural consequences. The third addresses how the mechanism can promote
process values.

\subsection{Addressing Social Choice Problems}

At present DAOs typically operate through a series of votes on proposals. If
vote buying or other forms of bargaining with side payments are not possible,
then each voting approach, whether directly or through electing
representatives, can be conceptualized as a mechanism with nontransferable
utility.\footnote{Nontransferable utility implies that players cannot bargain
with each other using money or some similar indicator of value to reach a
mutually agreeable result. Pareto improving moves, where for instance party A
buys off party B to achieve a particular outcome that results in gains for A
that outweigh the losses for B are not possible.} As such, a large series of
well-known potential ``social choice" pathologies arise. A striking and
relevant instance is the ``McKelvey-Schofield Chaos Theorem" derived by
\citet{McKelvey_1976} and \citet{Schofield_1978}. This Theorem states that if
the choice space is more than one-dimensional and preferences are Euclidean
(decline with distance from an ideal point) or, more generally, are convex as delineated in \citet{Schofield_1978}, then: (i) majority voting is unstable in the sense that every alternative is dominated by at least one other alternative; and, most strikingly, (ii) a series of majority votes between two alternatives can lead to \emph{any} alternative in the choice space, even ones that are Pareto dominated. Most DAOs operate through a sequence of such majority votes, and it is unlikely that the choice set of possible directions of change consistently reduces to one dimension that fully captures preferences.

Most generally, there is the Gibbard-Satterthwaite Theorem derived by
\citet{Gibbard_1973} and \citet{Satterthwaite_1975} which states that if
individual preference orderings are complete and transitive but otherwise
unrestricted and there are at least three alternatives, then a direct
mechanism is dominant strategy incentive-compatible if and only if it is
dictatorial.\footnote{ \citet{Borgers2015} provides a concise discussion and
proof of this Theorem and its significance.} In a rough sense what the
Theorem means is that with unrestricted preferences, a mechanism in which
each individual votes sincerely, recording their actual preferences, will
only work if there is a single individual who decides everything or if there
are no more than two alternatives. Once strategic voting enters the picture,
results can become unpredictable, difficult to estimate, and possibly very
unrepresentative compared to voters' actual preferences.\footnote{The
discussion by \citet{Tabarrok_1999} about how strategic voting might have
affected the U.S. Presidential election of 1860 under various voting regimes
is a good example.}

The mechanism allows a way around these social choice problems. The basic
auction creates a determinate outcome, one that awards control to the party
that claims it can perform in a way that has the most benefit for the other
token holders, a claim that is backed up by a value deposit. Initiating an
auction not only creates determinacy, but also is a way to address any
inferior operational or procedural outcomes that emerge from the social
choice process.

\subsection{Addressing Empty Voting} \label{empty}

In addition to the social choice difficulties with voting approaches, there
is another entirely separate set of potential problems associated with what
has been termed ``empty voting." Following the seminal work of
\citet{HuBlack2006}, we use the following terminology. \emph{Empty voting}
occurs when a party is able to exercise the voting rights of a token without
holding the associated economic rights to token value appreciation and any
distributions. \emph{Hidden ownership} is the opposite: the party holds the
economic rights without the right to vote the token, and, typically, without
appearing to be an owner in any corporate or blockchain register.

Empty votes can easily be created at little or no cost by a variety of means.
A party can borrow tokens and then vote them, leaving the economic ownership
to the lender. A party can engage in an equity swap, offloading the economic
rights and retaining the votes.\footnote{For example, the party starts with
some tokens then swaps the economic return from the tokens for the economic
returns of, say, a Treasury bond of equal value. The party still holds the
tokens and can vote them, but the economic interest is in the hands of the
swap counterparty.} There are many methods that employ
derivatives.\footnote{For example, the party holding a token can write a
call, buy a put, and borrow from a counterparty. The short call removes the
token upside, the long put eliminates the downside, and the party can lend
out the cash to pay the interest on the amount borrowed. The party is left
with no economic position at all, but the party still formally owns the token
and can vote it. The cost of entering this position will be nominal except
possibly for some fees, which will be low if there are active markets or if
potentially competing over-the-counter counterparties are readily available.}

Empty voting exists in conventional markets and has been documented by
\citet{HuBlack2006} as well as in a substantial literature following them.
There also are identifiable instances of empty voting in cryptocurrency
markets along with an awareness of the possible use of empty voting among
participants in those markets.\footnote{\citet{Buterin_2021} discusses ``vote
buying" and presents a theoretical example of empty voting that consists of
the equivalent of an equity swap. \citet{Copeland_2020} describes an actual
example from the takeover of Steem in 2020 by Justin Sun.} Commentators,
including \citet{HuBlack2006}, consistently point out that empty voting can
have positive as well as negative effects. \citet{Brav_2011}, for instance,
model whether empty voting is likely to have a net positive or negative
effect on corporate governance.

Whatever the balance between the positive and negative effects, empty voting
creates an element of arbitrariness because prevailing may be a matter of
more effectively accumulating empty votes rather than the result of a being
able to create more value or virtue.\footnote{As \citet[p. 907]{HuBlack2006}
state in the corporate context, potential use of empty voting leads to a
situation in which:
\begin{quote}
Voting outcomes might be decided by hidden warfare among company insiders
and major investors, each employing financial technology to acquire votes.
Adroitness in such financial technology may increasingly supplant the role
of merit in determining the control of corporations.
\end{quote}}

In the context of a battle for control, the mechanism developed here makes
empty voting irrelevant, avoiding any possible accompanying arbitrariness.
Any party can initiate an auction, and to win control, a party must submit
the best bid, one that promises an outcome that delivers the largest possible
social surplus to the other token holders, with a guarantee in the form of a value
deposit. It does not matter how many conventional token votes the party or
its competitors have, empty or otherwise.

Empty voting can facilitate value destruction. A party can take a large empty
voting position, combine that position with a net negative economic interest
in the token such as a collection of put options, and then vote for proposals
that reduce token value.\footnote{There may be other motives to destroy the
token such as being a marketplace competitor. \citet{huhamermesh2023} discuss
a general class of ``related non-host assets'' situations where an investor
might use empty votes to damage the value of one entity in order to enhance
the value of another entity in which the investor holds a substantial equity
stake.} In the most extreme case, the party could promote governance
decisions that totally destroy the value of the token and the DAO project.
The mechanism proposed here creates potential protection against such value
destruction strategies because it offers a profitable auction route that
renders empty voting ineffective.

The mechanism itself implicitly relies on empty voting because it permits a
party winning control to prevail in all votes during the ensuing control
period despite falling short of holding the required majority of tokens. As a
consequence, potential value destruction by the control party is a concern.
As discussed in the Introduction and described in detail in section
\ref{mechanism}, the mechanism contains measures to deter value destruction
by a control party regardless of how it arises. Value destruction triggers a
potential loss of deposits that outweighs any potential benefits and that
compensates existing token holders for any resulting drop in token value.

Designing the mechanism involves considering possible deleterious use of
hidden ownership as well as empty voting. In particular, subsection
\ref{toehold_danger} describes the need for accurate toehold reporting
accompanying auction bids. Hidden ownership is one way to conceal part or all
of the bidder's token position, and the mechanism must defeat use of such a
device even if it is hard or impossible to detect as such.

More generally, it is important for any DAO governance approach or mechanism to be \emph{EV-robust} in the sense of being effectively resistant both to deleterious uses of empty voting or hidden ownership and to any tendency for the decision mechanism to be compromised or blurred as a result of either manipulation.\footnote{The term ``EV-robust" uses the initials ``EV" to stand in for ``empty voting." Whenever hidden ownership is created, it is necessarily the case that there will be an offsetting empty voting position. Thus, ``EV-robust" is an appropriate descriptor that can refer to the hidden ownership side of the pairing as well to empty voting itself.} This task is complicated by the fact that empty voting and hidden ownership are easy to conceal, particularly because the elements that result in the empty voting or hidden ownership positions may otherwise have a legitimate hedging or other purpose. The mechanism will not be EV-robust unless it is impervious to concealed empty voting or hidden ownership.

\subsection{Promoting Process Values}

Although voting approaches have serious potential flaws, voting mechanisms
and other governance elements may have a \emph{process value} to participants
independent of operational efficiency. One would expect that process value
would be captured in token value because token holders will be willing to pay
more to participate in a DAO with governance features they value. Three
implications follow with respect to the auction mechanism.

First there is \emph{rescue}. If the indeterminacies and potential
pathologies of voting threaten the coherence or direction of the DAO during
an open period, causing the value of the DAO to drop, an auction that
initiates a control period is a remedy. The control party can set the DAO
back on course during the control period and then reinstate the voting
mechanism after addressing the threats. This feature may create a safe zone
of operation for voting mechanisms that serve important participation or
other normative goals despite their potential
flaws.\footnote{\citet{Hall_2022} describes elegantly how the history of
democracy can inform the design of DAOs. Among other elements, he considers
the various forms that delegation can take, including delegates of the token
holders appointing ``managers ... who could take certain operational
decisions more expertly than tokenholder voting could," but who remain
``accountable to tokenholders because they can be fired by the elected
delegates at any time." He notes that this approach bears ``similarities to
corporate governance." The mechanism here can be seen as an extension that
adds a way to right the ship through temporary, contestable delegation of
control to a competent party if the usual voting and delegation mechanisms
break down. It is analogous to external corporate governance through the
market for corporate control.} Otherwise, the voting mechanism may not be viable as a long-term way to operate the DAO.

Second, the auction mechanism is a means to promote process values, including
various desired voting approaches, by facilitating innovations in the default
governance state. If an increase in token value is attainable by shifting the
governance mechanism in a way that increases its process value without a
fully offsetting loss in operational efficiency, then there is the potential
for an auction to create the shift. The sequential aspect of the auction
mechanism is designed to achieve this result. Successful implementation of a
shift in the default governance mechanism after winning an auction will raise
token value, ending the control period with a reversion to the new superior
default governance mechanism.\footnote{The ideal situation is one in which
the winning bidder sets the token target price just high enough to fully
reflect the increase in token value from the shift in default governance
mechanism. Execution results in a rapid if not instantaneous increase in
token value to the target level, which ends the control period. If the
winning bidder set the token target price too low, then the same result
occurs. If the winning bidder sets the token target price too high, then, as
discussed in section \ref{mechanism}, the existing token holders will enjoy
any overage at the expense of the winning bidder.} It may be difficult or
impossible to achieve a possibly complex innovation in the default governance
state through voting or other procedures that comprise the current state.

Third, the fact that process value is reflected in token value means that a
DAO that has sacrificed operational efficiency to add an even larger amount
of process value will be immune from a takeover through the auction mechanism
that eliminates the process value to increase operational efficiency. A party
intending to implement this move through an auction would be unable to
initiate the auction if the DAO is currently fully valued and would lose in
the auction if the DAO is currently undervalued.\footnote{Initiating an
auction requires a bid with a token target price in excess of the current
token value. If the DAO is undervalued but the current operational efficiency
plus governance mechanism maximize the total value of the DAO, then a bidder
maintaining the status quo will win the auction. See section \ref{mechanism}
for details.}

\section{Conceptual Overview} \label{roots}

We assume the subject DAO is an economic DAO. We consider a setting similar
to \citet{Burkart2021} in which various parties can use costly effort to
increase the value of the DAO. In particular, suppose each party can engage
in various projects that consist of expending effort, labor, and resources
equivalent to $C$ monetary units, in order to increase the value of the DAO
by $N(C)$ monetary units.\footnote{These costs are external to the DAO and are borne by the party engaging in the project. If the party uses DAO resources such as DAO treasury assets to execute the project, use of these resources will diminish the value of the DAO directly, reducing $N(C)$ rather than being an addition to $C$.} Suppose that the DAO has $q$ tokens outstanding and
that the current market price per token is $P_0$. Define $V = V(C) = N(C)/q
+P_0$ to be the token value emerging from a particular business plan. This
business plan will create \emph{social surplus} $\psi=N(C)-C=(V-P_0) q -C$. Consider that
expectations about the nature of future value-additive projects and about the
distribution of social surplus from those projects will affect the initial funding
value of the DAO, possibly being critical to having enough funding to start
up at all. In other words, the expected later treatment and facilitation of
value-added projects will have an \emph{investment impact} on initial funding
for DAOs, and also, in an obvious way, on the on-going investment value of
the DAO, which continues to depend on possible future projects.

The target is the following first-best outcome:
\begin{enumerate} [label=\Alph*)]
\item At any point after initiation of the DAO, the mechanism will
    facilitate implementation of the value-additive project that results
    in the highest social surplus $\psi$.
\item When such project is created and implemented, the mechanism will
    allow the creating party to cover its cost but will allocate the
    social surplus entirely to the other token holders.
\end{enumerate}
If both targets are met, the result will be the highest positive investment
impact as initial and on-going investors will receive the maximum possible
benefit from future innovations.\footnote{\citet{Posner_2017} envision a system of ``partial common ownership" in which there are property taxes and contestable control, at least periodically, for various kinds of property through a Harberger tax mechanism. Central to a Harberger tax is a method of assessment aimed at eliciting the owner's actual valuation. The owner states a value that will be the basis for imposing a property tax, but understatement is policed by the right of the state or others to buy the property at the claimed value. In the case of multiple potential buyers, something like the mechanism here is required to allocate the property to the buyer who can produce the highest possible social surplus, the owner's valuation being equivalent to $P_0$ in the model here.} In addition, at each point in time the DAO
will implement the on-going business plan that adds the highest possible
amount of social surplus.

To aim at the target, we create a sequential auction mechanism that produces
periods of temporary contestable control interspersed with periods in which
the DAO reverts back to a default governance state. The mechanism
has the property that the highest social surplus project is chosen subject to
some constraints that guarantee the continuation of market trading. Some
social surplus necessarily leaks to the project creators, resulting in an
outcome that falls short of first best.

To describe the setting further, we use some terminology from
\citet{Burkart2021}: \emph{Jensen-Meckling free riders} and
\emph{Grossman-Hart free riders}.\footnote{The names derive from phenomena
described in \citet{Jensen_1976} and \citet{Grossman_1980a}.}
\citet{Jensen_1976} point out that when the party who manages a corporation,
expending all the effort, owns less than all of the common stock, the costly
effort is matched with only part of the gains. The other shareholders are
Jensen-Meckling free riders, reaping gains without bearing any of the costs.
These free riders, however, are the successors of the original investors or
of parties who contributed effort previously. A policy of rewarding them
encourages initial or subsequent investments in the corporate project, having
the investment impact described above.

The conventional picture of Grossman-Hart free riders is a corporate
enterprise for which the equity holders consist of a large group of parties
all of whom own a very small stake. The chance that any one such party will
be decisive in a vote is minuscule, which, combined with the large number of
holders, creates a collective action problem. Consider a project creator who
can profit by building up a share ownership position and then announcing or
implementing a project that increases the value of the equity. If the equity
holders get wind of the project, they will free ride by refusing to sell at
less than the post-project target value of the equity. Open market purchases
by the creator will push the price up, and in the U.S., the creator will have
to reveal its holdings and intentions once the holdings exceed 5\% of the
total equity. The free-riding by the equity holders will limit the portion of the total added token value
that the project creator can extract, potentially killing the project if the
extractable portion of the total added token value is lower than the creator's costs. Alternatively,
creators will pick projects that do not maximize social surplus but are
viable based on the increase in market value available from a modest
toehold.

\citet{Burkart2021} create a model based on the value impact of effort that
captures the current situation for project initiation through activism and
tender offers in the United States. The tender offer route is restricted by
Grossman-Hart free riding after an offer is made, limiting the offeror to the token value
surplus from a toehold. Activists proceed through a costly campaign aimed at
managers and other shareholders to initiate a new project, avoiding
Grossman-Hart free riding, but still being subject to Jensen-Meckling free
riders. Revenues from activism are again limited to a toehold, but the other
shareholders potentially benefit from the activism without bearing any costs.
\citet{Burkart2021} also analyze a third route: activism directed at
initiating a merger, which they term ``takeover activism." In a merger, the
price that shareholders receive is set by the managers of the two firms,
allowing additional total added token value to be made available by forcing the dispersed shareholders to
accept a price below the target price.\footnote{Most states require
shareholders to approve a merger by a vote. It will be in the interest of
shareholders to do so despite not receiving the target price if a lower price
is necessary to make the merger work by providing enough surplus for the
acquiring party, here the project creator. Grossman and Hart free riding
otherwise creates a collective action problem that potentially precludes
shareholder gains entirely.} \citet{Burkart2021} survey the empirical
evidence and note that among activist projects, the high-return ones for
shareholders are concentrated among instances of takeover activism.

The mechanism created here addresses the potential value-reducing free-riding of both types through two devices. First, by creating a \emph{freeze-out} feature as part of an auction, the mechanism addresses the Grossman-Hart free-rider problem by allowing the project creator to buy a proportion of the other token holders' positions for no premium. Those token holders will earn the full amount of the expected total added token value on the retained proportion that is not purchased by the creator. This feature makes the auction equivalent to a merger in which token holders receive only a portion of the expected total added token value on their token position shares because of the terms of the merger which dictate the proportional stake in the merged entity that they receive in exchange for their shares in one of the precursor entities.\footnote{Although the mechanism creates an outcome somewhat analogous to a merger, no actual merger is involved. Trading in the DAO remains continuous, and the life of the DAO goes forward. A merger in which the acquiring party buys all of the DAO shares would mean the end of the DAO in its current form. It might be that the acquiring party, which may be a shell, is itself set up to be a DAO, perhaps one with a new set of smart contracts meant to upgrade the acquired DAO. We leave exploration of this possibility to future work, including the case in which the selling token holders receive tokens in the new DAO instead of a cash-equivalent in exchange for the tokens in the old DAO.} The bidding mechanism creates an incentive for bidders to limit the amount of total added token value that they attempt to claim via the freeze-out feature to the lowest possible value, the amount required to cover their costs taking into account the expected gain on their toehold position. As discussed in the next section, there typically will be some leakage because the winning bidder will have some scope to go beyond that limit.

Second, there is the problem of post-auction incentives in the face of
Jensen-Meckling free riding. At the end of the auction, the winning bidder has
control but owns less than all of the tokens. The auction mechanism is
designed to allocate social surplus to the other token holders by limiting the stake
held by the winning bidder, and that stake may be much less than half of the
outstanding tokens. The mechanism restores full incentives to execute through
the value deposit described in the Introduction and discussed more fully
below. To the extent that a control party falls short of the target price,
they have to cover the shortfall for all of the other token holders in addition to losing out on their own token position in the DAO. As a result, they have an incentive to execute that is at least as large as a party that has 100\% ownership, eliminating any adverse Jensen-Meckling impact on post-auction execution incentives.\footnote{Complexities arise if control parties hedge part or all of their token position. Subsection \ref{EV_robust} discusses this possibility and potential responsive adjustments to the value deposit and the function specifying value deposit forfeit amounts.}

As discussed in subsection \ref{basic}, the value deposit and also the surety
deposit employed by the mechanism are equivalent to forcing the control party
to take on option positions. Other researchers, most prominently,
\citet{Burkart2015}, have described the strong potential role of requiring
such positions in implementing a cogent market for corporate
control.\footnote{\citet{Burkart2015} show that a signaling equilibrium in a
setting in which there are private benefits to control can attain the full
information outcome by combining a cash offer with an offer to sell a call
option with an exercise price equal to the bid price. Here the required
option positions implicit in the deposits serve multiple roles, including one related to signaling.} We use
deposits rather than requiring option positions because deposits are code
feasible, while derivatives require counterparties which raises issues of
trust.

The approach here flows from a very general deposit-based model: There is a single auction deposit, $D_a$, and a deposit forfeit function $\Phi(D_a,P_0,S,X)$ where $P_0$ is the token price when the auction is initiated, $S$ is the value per token that the winning bidder claims is attainable, and $X$ is an outcome, typically the token price when the winning bidder's control terminates. The first three parameters in the function are set by the time the auction ends and the winning bidder takes control. The deposit forfeit function specifies the amount of the deposit that is forfeited for each possible value of $X$ given the three auction parameters. The function is not required to be continuous in $X$, and we will see that discontinuities can play an important role in creating incentives for control parties. 

Instead of operating at the level of full generality by choosing the deposit amount and the deposit forfeit function to optimize some set of objectives, we consider a more limited version that employs two deposits and certain restrictions on the deposit forfeit function. This approach allows us to illustrate the features of the auction mechanism more clearly. We leave a more general treatment to future work.

The goal throughout is to create an example of a mechanism that has some
plausibility and likely effectiveness in order to introduce the idea of
auction-based temporary contested control for governing DAOs. No claim is
made that the mechanism is optimal among the set of all such mechanisms. We
discuss possible alternative features at many points, not attempting to come
to a conclusion concerning whether they are superior or inferior to the main
variant that we describe.

With this overview in hand, we present the mechanism formally and evaluate it
in the next two sections along with two appendices. Section \ref{mechanism}
contains three subsections, describing respectively, a single auction,
post-auction operation, and subsequent auctions, along with some evaluation
of the mechanism. Section \ref{more} completes the task of evaluation,
organized topically in separate subsections. One appendix consists of proofs.
The other appendix explores the consequences of adding certain stochastic
elements to the model developed in the text.

\section{A Sequential Auction Mechanism} \label{mechanism}

We construct a \emph{sequential auction} consisting of a series of
\emph{basic auctions}, implemented via one or more smart contracts,
collectively ``the DAO Code." The DAO Code enables the DAO to operate through periodic basic auctions, with a fixed \emph{control period} between auctions, subject to early
termination under some circumstances. The DAO Code also permits a basic
auction to be initiated by any party at any time, independent of the periodic
auctions. These features make the control created by the auction both
\emph{temporary} and \emph{continuously contestable}.

We develop the sequential auction mechanism in a deterministic setting
in which the token price moves only based on market assessments of the progress and prospects of existing and potential future business plans. In Appendix A.2 we consider what might happen in a stochastic setting, examining possible adjustments to the mechanism and possible hedging by winning auction bidders.\footnote{The stochastic elements discussed in Appendix A.2 include sources of value fluctuations such as the impact of broader market movements on the DAO token value that are not necessarily related to the cogency of business plans or the quality of plan execution.}

\subsection{The Basic Auction}

\subsubsection{The Basic Auction Mechanism} \label{basic}

A basic auction is initiated at a time $T_0$ by a first bidder making a bid.
The basic auction is open for the competitive bidding process up until a
winner is determined at some later time $T_1$, with the fixed total time
length $T_{auction} = T_1-T_0$ of the bidding period specified in advance by
the DAO Code. Suppose that at $T_0$ the price of the DAO governance token is
$P_0$ and $q$ tokens are outstanding. A bid, $\beta = (S, R, t_b)$, consists
of three parameters: a \emph{value claim}, $S \geq P_0$, per token; $t_b$,
the proportion of the q tokens held by the bidder at $T_0$, hereinafter
termed the \emph{bidder's toehold}; and a \emph{surplus claim}, $R \leq
(t_m-t_b)(S-P_0)q$ where $t_m$ is a parameter set by the DAO Code that, as discussed in subsection \ref{efficiency}, plays a role in assuring adequate market liquidity. The
bidder's toehold, $t_b q$, is deposited into the applicable smart contract
simultaneously with the bid, becoming part of the bidder's required
\emph{token deposit} under the mechanism. The basic auction is an English
auction, ascending in the \emph{auction parameter} $A = (1-t_b)(S-P_0) q-R$.
All prices and quantities such as $S,~ P_0, ~\mbox{and}~ R$ are denominated
in the units of a particular \emph{reference fiat currency} designated by the
DAO Code.

From this point on, we assume for convenience that the total number of tokens
remains at $q$. Then token \emph{price} differences translate linearly into
\emph{total value} differences through the multiplicative factor $q$ even if
the prices are realized at different times. This assumption avoids having to
continually correct for possible changes in the number of tokens outstanding,
which is trivial but cumbersome.

A central feature of the basic auction is a \emph{freeze-out} element which allows the winning bidder to force the other holders as of time $T_1$ to sell $t_f q$ of their tokens to the bidder at price $P_0$. The main function of the surplus claim $R$ is to determine $t_f$, in particular, $t_f q = \frac{R}{(S-P_0)}$. If the value claim $S$ ends up being correct, then the forced sale results in a transfer of gain equal to $R = t_f (S-P_0)q$ from the other holders to the bidder, the basis for the terminology ``surplus claim." $t_f$ is the \emph{freeze-out proportion}. $R < 0$ is possible, in which case $t_f < 0$, and the bidder will offer $\left| t_f \right| q$ tokens at $P_0$ to the other holders subject to the total holdings of the bidder, i.e., $\left| t_f \right| \leq t_b$. At the end of the forced sale or purchase of tokens, the bidder holds the proportion $t_d = t_f + t_b$ of the total tokens, and all of these tokens must be deposited in the applicable smart contract, resulting in the token deposit at the end of the auction totaling $t_d q$.\footnote{The bidder is not barred from buying tokens during the auction, that is between time $T_0$ and time $T_1$. Any such token purchases do not need to be reported or added to the token deposit. It is likely, and we are assuming, that any such buying after the announcement of the bid will be futile both during the auction period and afterwards because Grossman-Hart free-riding effects as well as market maker reactions will defeat the bidder's attempt to secure more of the potential added token value. The toehold position is discussed further in subsections \ref{toehold_danger} and \ref{toehold_role} infra.}

The intuition behind the auction parameter, which will emerge with a more rigorous meaning in the Propositions below, is the following. $S$ represents a claim by the bidder that the project will reach value $Sq$ after the bidder gains control. The auction parameter is $A = (1-t_b)(S-P_0) q-R = (1-t_b-t_f)(S-P_0) q =  (1-t_d)(S-P_0) q$, which is precisely the portion of the \emph{total added token value}, $(S-P_0)q$, that the bidder will deliver to the other token holders if $S$, the bidder's value claim, is realized. If the token value does not reach $S$, then as discussed below, transfer of part or all of the value deposit to the other token holders will ensure that they are at least as well off as if the value $S$ was realized. These other token holders will receive at least their promised portion of the total added token value whether or not the winning bidder successfully executes the project.\footnote{This feature is an important aspect of the attempt to maximize the investment impact of the mechanism despite the fact that most of the non-bidding token holders are likely to be Jensen-Meckling free riders for whom assessing alternative business plans would not be rational given the size of their stakes. If the winning bidder makes an excessive value claim that results in the defeat of a more capable bidder in the auction, the  non-bidding token holders are protected automatically at the expense of the winning bidder who failed to perform as claimed.}

The DAO Code imposes a \emph{market size condition}, $t_d = t_f + t_b \leq
t_m$. This condition guarantees that a minimum proportion $1-t_m$ of the
total tokens remain in the market during and at the end of the auction, which
may be necessary to ensure a functioning market after the auction is
over.\footnote{The freeze-out step is proportional, leaving each holder
immediately after the auction ends with the proportion
$\frac{1-t_f-t_b}{1-t_b}$ of their time $T_1$ holdings. I.e., the same time
$T_1$ market participants have holdings, albeit proportionately reduced. This
feature should allow the market to continue to function smoothly across the
transition point.} Choosing any $t_m < 1$ limits the portion of the total added token value that can be shifted to the winning bidder to an amount less than the total added token value
available.\footnote{Under the mechanism, $t_m$ is the maximum proportion of tokens that the control party can hold as part of the token deposit, leaving the rest to be freely traded. I.e, $t_d \le t_m$. The subscript ``m" is chosen to signify ``maximum," while the subscript ``d" signifies ``deposit."}   If the bidder has high enough costs, this limitation may cause a
surplus-producing business plan not to be viable under the mechanism.

In addition to the \emph{token deposit} of the $t_d q$ tokens, the bidder is
required to make three additional deposits in the form of stablecoins of
types permissible under the DAO Code representing units of the reference fiat
currency:

\begin{enumerate}[label=\arabic*)]
\item A \emph{value deposit}: $D_v > (1-t_d)(S-P_0) q$.
\item A \emph{purchase deposit}: $D_p = \max \left\{ \frac{R}{S-P_0} P_0,
    0 \right\}$.
\item A \emph{surety deposit}: $D_s = \max \left\{(1-\gamma)P_0q -
 D_v, 0\right\}$, where $\gamma$ is defined and
    discussed below.
\end{enumerate}

The bidder is making a value claim, $S$, that the bidder's business plan for the DAO will suffice to increase its value to at least $S q$. The DAO Code may refund part or all of the value deposit when the bidder's control comes to an end, with the rest forfeited and paid to the other token holders. A \emph{value deposit forfeit function}, $\phi(D_v,P_0,S,P_{ref})$, specifies for each value of a \emph{reference price}, $P_{ref}$, the \emph{value deposit forfeit amount} given the three auction parameters $D_v$, $P_0$, and $S$. The amount of the value deposit that is refunded is $D_v - \phi(D_v,P_0,S,P_{ref})$. When the control period is not followed immediately by a subsequent auction, the value deposit forfeit amount is precisely $\phi$. In this situation, as discussed below, the forfeited amount is transferred to the token holders of record other than the winning bidder as of the end of the auction.

Consider a \emph{standard case} of a value deposit, reference price, and
value deposit forfeit function defined by:
\begin{enumerate}
\item $D_v = (1-t_d)(S-P_0) q (1 + \epsilon)$ where $\epsilon > 0$ is
    small or infinitesimal, creating only a slight deviation from 1.
\item $P_{ref} = P_{end}$ where $P_{end}$ is the token price at the end
    of a control period that is not immediately followed by a subsequent
    auction.
\item $\phi(D_v,P_0,S,P_{end}) = \max \left\{ 0, \min \left\{D_v,(1-t_d)(S-P_{end}) q (1 + \epsilon) \right\} \right\}.$
\end{enumerate}

\noindent In this standard case, a winning bidder who falls short of attaining a token price equal to the value claim will forfeit to the other token holders an amount just a tiny bit more than the additional gain that the other token holders would have realized if the token price had reached $S$ instead of $P_{end} < S$.\footnote{An additional property of the standard case is that the value deposit combined with the value deposit forfeit function can be conceptualized as requiring the bidder to write an in-the-money bear put spread in favor of the other token holders consisting of being long a put with strike price $S$ and short a put at strike price $P_0 < S$. \label{options note}}

Some aspects of the standard case generalize. Under the mechanism, the value
deposit forfeit function is always set in such a manner that the token holders
other than the bidder will receive \emph{at least} the full gain $S - P_0$ per token
on their $(1-t_d)$ share of the tokens whether or not the business plan is
successfully implemented. If the business plan succeeds, the other token holders realize the full gain because the market price increases to at least $S$, fulfilling the value claim. If not, operation
of the value deposit forfeit function ensures that any deficit is more than made up from the
value deposit.\footnote{In cases in which the bidder's control ends in a
subsequent auction, implementation of any refund may be delayed until after
subsequent control periods and may become moot in whole or in part due to the
performance of the token price under the guidance of the subsequent control
parties. To the extent the refund is not paid out as part of the transition
to the next control period, the mechanism preserves it as an obligation. See
subsections \ref{post_auction} and \ref{subsequent_auctions} infra.} Thus, the
value deposit combined with the operation of the value deposit forfeit function transforms the bidder's value claim into a commitment to deliver the promised increase in token value to the other token holders. The fact that this deposit is held by the
applicable smart contract makes this commitment credible and immediately
enforceable in a code feasible manner.

We stated the value deposit as an inequality above, $D_v > (1-t_d)(S-P_0) q$, rather than as a specific amount. Because the value deposit plays multiple roles, it is valuable to leave flexibility because the optimal amount of the value deposit along with the best accompanying value deposit forfeit function may be application specific. We have already mentioned a role that is satisfied by a value deposit close to the $(1-t_d)(S-P_0)q$ bound: guaranteeing the transfer of the amount of gain that the winning bidder has promised to token holders other than the bidder, which helps attain the goal of shifting the maximum possible amount of social surplus to these other token holders. But other considerations, several of which we discuss in ensuing subsections, suggest that a value deposit significantly larger than the bound may be useful. 

The reason there is a lower bound, and the particular stated one, relates to another role of the value deposit: ensuring that the bidding mechanism works successfully. One goal of the mechanism is to induce the bidder to make a value claim equal to the bidder's honest assessment of the token value attainable under the envisioned business plan. Consider the following Lemma and Corollary of the Lemma, proven in Appendix A.1. The context is a bidder who makes a value claim of $S$ and faces cost $C$ to implement a business plan that the bidder believes will increase the token value to $V$ from the price $P_0$ prevailing at the initiation of the auction.\\

\textbf{Lemma 1.} \emph{Given a business plan $(V,C)$, a bidder will avoid making a value claim $S > V$ if and only if for every possible outcome $X \in [P_0, S]$, $\phi(D_v,S, P_0, P_{ref} = X) > (1-t_d)(S-X)q$ under the value deposit forfeit function.}\\

\noindent In words, for any outcome $X\in [P_0,S]$, the value deposit forfeit function must distribute more than $(1-t_d)(S-X)q$ of the value deposit, $D_v$, to the $T_1$ token holders rather than refunding it to the control party.\footnote{Note that $V \in [P_0,S]$ is private information known only to the bidder. The mechanism designer must specify a value deposit forfeit function that achieves the desired result for the case $S>V$ regardless of where $V$ falls within $[P_0,S]$.}  Defining the added amount as $\delta(S,X)$, note that the all the Lemma requires is that this amount be positive for any $X \in [P_0,S]$. Although the Lemma restricts the value deposit forfeit function, the restriction leaves considerable latitude. For instance, the mechanism designer can set $D_v$ at a much larger value than $(1-t_d)(S-P_0)q$, say twice as large, and then impose high additional forfeitures versus $(1-t_d)(S-X)q$ only in certain ranges within $[P_0,S]$.\footnote{Because the value deposit plays multiple roles, the freedom to set $\delta(S,X) > 0$ at will must be applied carefully. For example, if the values of $\delta(S,X)$ are set in such a way that the quantity $(1-t_d)(S-X)q + \delta(S,X)$ does not monotonically decline over the entire range $[P_0,S]$, then there will be at least subranges in which the control party has no incentive to make efforts to increase the value of the DAO. See subsection \ref{EV_robust} infra (discussing post-auction incentives).}

Setting $X=P_0$, a corollary follows immediately from the fact that
$D_v \ge \phi(D_v,S, P_0, P_0)$:\\

\textbf{Corollary 1.} If the restriction on the value deposit refund stated in Lemma 1 is met, \emph{$D_v > (1-t_d)(S-P_0) q$}.\\

\noindent This Corollary is the source of the inequality we have used to define the value deposit above.

We now state the value deposit forfeit function formally and identify an important special case that is useful as a baseline in what follows:
\begin{equation} \nonumber
    \phi(D_v,P_0,S,X) =
        \begin{cases}
                0 & \text{if}\ X > S\\
                (1-t_d)(S-X)q + \delta(S,X) & \text{if}\ X \in [P_0,S]\\
                D_v & \text{if}\ X < P_0 
        \end{cases}
 \end{equation}
\noindent where $\delta(S,X) > 0$. Define a \emph{standard value deposit forfeit function} as one for which $\delta(S,X)$ is infinitesimal for all values of $S$ and $X$. Then for $X \in [P_0,S]$ we have $\phi(D_v,P_0,S,X) \approxeq (1-t_d)(S-X)q$.\footnote{The standard value deposit forfeit function is equivalent to the value deposit forfeit function used to define the standard case above.}

Note that if $D_v > (1-t_d)(S-P_0)q$ by more than an infinitesimal amount, then there is a discontinuity in the standard value deposit forfeit function at $P_0$. Define $\phi_0 = D_v - (1-t_d)(S-P_0)q$ to be the \emph{baseline loss penalty}. As the name suggests, the value deposit forfeit amount jumps by $\phi_0$ when the token value falls below $P_0$ and there is a loss relative to the token price when the auction was initiated. Any outcome less than $P_0$ will be penalized by at least this amount. As discussed in subsection \ref{value destruction}, this baseline loss penalty can be very useful in deterring value destruction. Now we discuss the other deposits and the operation of the auction.

The purchase deposit ensures performance when a bid with $R>0$ commits the
bidder to buy $t_f q$ tokens at price $P_0$.

The surety deposit addresses the danger that the bidder will engage in value
destruction after gaining control of the DAO. Note that the surety deposit is
reduced to the extent of the value deposit. As will become apparent, these
two deposits working in conjunction perform three functions: creating optimal
bidding incentives, incentivizing performance of the business plan by the
winning bidder after the auction ends, and deterring value destruction. The
choice of the levels of the deposits and the applicable forfeit conditions for
each deposit reflect the confluence of these three goals. We discuss the
overall role of the surety deposit including the choice and significance of
the parameter $\gamma$ in subsection \ref{value destruction} after describing
the auction mechanism further. Choosing optimal levels for the two deposits also depends on stochastic elements discussed in Appendix A.2.

Consistent with the lack of centralized management in a DAO, the auction is
designed to be self-executing through the Auction Contracts. To reach that
goal, the auction mechanism must be code feasible. All four deposits are
useful in that respect. Because of the purchase deposit requirement,
non-payment cannot derail the auction. There is no need to have recourse
outside of the Code to legal process for purposes of collection. As noted in
the margin,\footnote{See note \ref{options note} supra, which states the equivalent derivatives position for a case in which that position is a simple options spread. In general, there will be an equivalent derivatives position, but it may not be simple.} the value deposit, combined with the value deposit forfeit function,
serves a function that also could be accomplished through derivatives. Use of
a deposit strategy eliminates the need for counterparties and the possible need to
enforce counterparty compliance with the option contracts, elements that may
not be code feasible without a great deal of added complexity or at all.
Similarly, the token and surety deposits substitute for mechanisms that would
rely on conventional derivative contracts and escrow arrangements enforced
through the legal system.

All four deposits as well as a valid bid are required for the bidder to
participate. If the bidder wins the auction, then the DAO Code uses the
purchase deposit to acquire the freeze-out proportion of tokens for transfer
to the winning bidder and retains the other three deposits, returning them if
and only if certain conditions are met. If the bidder loses the auction, all
four deposits are returned.

The auction is closed at time $T_1$, a date which is analogous to a record
date in corporate stock transactions, and the following \emph{Auction Closing
Steps} are implemented instantaneously:\footnote{Blockchain technology allows
for continuous identification of token holdings, and the steps outlined below
can be implemented through smart contracts instantly with respect to the
token holdings as of time $T_1$ when the auction closes.}

\begin{enumerate} [label=\arabic*)]
\item The DAO Code identifies token holders of the $(1-t_b) q$ total
    tokens other than the bidder's token deposit as of time $T_1$,
    collectively, the \emph{$T_1$ token holders} who hold the \emph{$T_1$
    token holdings}.
\item If $t_f \geq 0$, the DAO Code transfers a total of $t_f q$ tokens
    from the $T_1$ token holdings on a pro rata basis to escrow in the
    appropriate smart contract as part of the bidder's token deposit,
    using the purchase deposit to pay each $T_1$ token holder $P_0$ per
    token transferred.
\item If $t_f < 0$, the bidder has chosen to reduce its holdings from the
    baseline token deposit, $t_b q$, by selling $|t_f| q$ tokens at price
    $P_0$, where $|t_f| \leq t_b$. The DAO Code initiates a sales process
    at $T_1$ when the auction ends, making the offering at price $P_0$
    first to the $T_1$ token holders and then to market maker external
    agents (human or automated smart contracts), if any, operating under
    the DAO protocol. Any proceeds are remitted to the bidder. If tokens
    remain unsold, they remain under the bidder's ownership as part of
    the token deposit. At the end of this process, the total token
    deposit is $t_d q$ where $t_d =t_b+t_f \geq 0$, assuming all the
    tokens are sold.
\item The DAO Code creates a \emph{dynamic vote pool} consisting of
    $v_T(q_T, t_d q)$ additional voting rights assigned to the bidder, an
    amount that is adjusted continuously based on $q_T$, the number of     
    tokens outstanding and eligible to vote at each future time $T$, to
    ensure that the bidder retains majority control of the DAO. The
    required condition is $v_T + t_d q > 0.5 (q_T + v_T)$, which can be met
    by setting $v_T = q_T -2 t_d q + 1$ where $q$ is the number of tokens 
    outstanding and eligible to vote as of time $T_1$ when the auction 
    ends. The votes in the pool are empty votes because they are not 
    matched with the corresponding economic interest inherent in the 
    tokens.\footnote{The choice of giving the winning bidder majority
    control ($> .5$ of the votes) implicit in this arrangement leaves open
    the possibility that
    the winning bidder will not be able to prevail on DAO issues that
    require a threshold greater than 50\% to pass the applicable
    proposal. The exact percentage established by the DAO Code for the
    vote pool needs to be sensitive to the hierarchy of voting thresholds
    and the associated issues set for the DAO more generally. If a less
    conventional choice method such as quadratic voting is employed by
    the DAO, the dynamic voting pool must grant whatever number of
    additional voting rights is required to establish control.}
    
\end{enumerate}

\subsubsection{The Optimal Business Plan and Bidding Strategy} \label{optimal_bid}

Suppose a potential bidder can implement a business plan $(V,C)$ that
involves expending effort, consisting of labor and resources equivalent to
$C$ monetary units, that will result in a token value $V > P_0$. This
business plan creates social surplus $\psi(V,C) = (V-P_0)q - C$. Suppose that
among all of the potential bidder's possible business plans, the business plan
$(V^*, C^*)$ creates the largest amount of social surplus, $\psi^* =
(V^*-P_0)q - C^*$.\footnote{\label{private_values} We assume that the auction
takes place in a ``private values" setting with asymmetric information. Each
bidder's potential set of business plans and their value-enhancing potential
are known only to the bidder. All bidders agree on the value of the current
operation. If there is a common values element in which bidders have
different signals concerning the value of the current operation and bid based
on perceived undervaluation at price $P_0$, then the efficiency and
surplus-distribution characteristics of the auction mechanism may be
affected. However, choosing the English auction as a mechanism tends to make
any such impacts benign or even beneficial. See note \ref{common_values}
infra.}

The potential bidder must choose both a business plan (V,C) and a bid $\beta
(S, R, t_b)$. It will turn out that the parameter $t_b$ is redundant with
respect to winning the auction. Only $S$ and $R$ matter. The following
Proposition characterizes the optimal business plan and bidding strategy:\\

\textbf{Proposition 1.} \emph{Suppose that the condition in Lemma 1 that precludes value claims with $S>V$ applies. Then the following is the optimal project choice and bidding strategy for a potential bidder:}\\

\noindent \emph{(i) Regardless of the bidding strategy chosen, the potential
bidder chooses the business plan $(V^*,C^*)$ that results in $\psi = \psi^*$,
the largest possible social surplus that the potential bidder can generate
subject to the market liquidity constraint, $t_d = t_f + t_b \leq t_m$. If
$\psi^* \leq 0$, the
potential bidder does not make a bid.}\\

\noindent \emph{(ii) The bid parameters that result in the strongest possible
bid are:}\\
\indent $S=V^*$ \emph{; and}\\
\indent $R = C^* - (V^*-P_0)t_b q$\\
\noindent \emph{which result in an auction parameter equal to $A^*$, the
largest possible social surplus that the potential bidder can generate subject to
the market
liquidity constraint:}\\
\indent $A^* = (V^* - P_0)q-C^* = \psi^*$.\\
\emph{This bid results in zero profit for the potential bidder, with all of the social surplus being shifted to the other token holders.}\\

\noindent \emph{(iii) If a profit level $\Pi_b$ is feasible given the market
liquidity constraint, then the strongest possible bid parameters are:}\\
\indent $S=V^*$ \emph{; and}\\
\indent $R = \Pi_b + C^* - (V^*-P_0)t_b q$\\
\noindent \emph{which result in the following auction parameter:}\\
\indent $A(\Pi_b) = (V^* - P_0)q-C^* -\Pi_b = \psi^* - \Pi_b.$\\
\emph{This bid results in profit equal to $\Pi_b$ for the potential bidder,
with the remaining social surplus, $\psi^*-\Pi_b $, shifted to the other token holders.}\\

\noindent \emph{(iv) The largest obtainable profit level is $\Pi_b = t_m (V^*
- P_0) q - C^* \leq \psi^*$ given $0 < t_m < 1$. The minimum social surplus that must be transferred to the other token holders is $F = (1-t_m)(V^* - P_0)q$.}\\

Leaving a formal proof to an Appendix, we outline the proof here with an
emphasis on intuition and then discuss the significance of the results.

The potential bidder's highest surplus business plan produces total added token value of
$N^* = (V^* - P_0)q$ at cost $C^*$. The potential bidder will realize the
proportion $t_d = t_f + t_b \leq t_m$ of this added value through a toehold of
$t_b q$ tokens acquired before the auction plus $t_f q$ tokens acquired at
$P_0$ from the $T_1$ token holders using the freeze-out feature of the
auction mechanism. The parameter $0 < t_m < 1$ guarantees that the $T_1$
token holders will receive at least the proportion $1-t_m$ of the total added token value, as
free riders. Under the auction mechanism, $t_f = \frac{R}{(S - P_0) q}$. For
any fixed level of $S$, choosing $R$ is equivalent to choosing $t_f$, a
residual that completely determines the division of the total added token value between the
control party and the free riders. For this reason, the fixed toehold
proportion $t_b$ is irrelevant to division of the total added token value, which depends
on $t_d = t_b + t_f$, where $t_f$ is freely chosen subject only to $-t_b \leq
t_f \leq t_m - t_b$.

Consider the auction parameter:
$$A = (1-t_b)(S-P_0)q- R = (1-t_f-t_b)(S-P_0) q.$$
It is clear that $A$ is increasing in $S$ holding $t_f$, and thus the
potential bidder's share of the total added token value, fixed. This fact tells us that the bidder will choose $S \ge V$. The remaining question is whether the bidder would ever opt for a value claim greater than $V$. 

To examine that question, consider the
potential bidder's profit function expressed in terms of $t_f$ when the
potential bidder implements a business plan $(V,C)$:

$$\Pi_b = (V -P_0)(t_f + t_b) q -C - \phi(D_v,P_0,S,V).$$

\noindent The first term is the added token value realized by the potential bidder,
the second term is the potential bidder's project cost, and the third term is
the expected value deposit forfeit that will result from choosing $S > V$ when the highest token value the bidder can achieve by executing the business plan is $V$. This expected value deposit forfeit requires a larger surplus claim, potentially weakening the bid. However, overbidding by choosing $S > V$ also increases the first term in the auction parameter, which admits a larger surplus claim. Resolving the impact of this tradeoff requires mathematical analysis. Lemma 1, set forth in the previous subsection, states restrictions on the value deposit forfeit function under which choosing $S>V$ means that earning the same amount of surplus requires a weaker overall bid versus choosing $S=V$.

The strongest possible bid subject to the value deposit forfeit function restrictions will minimize $R$ in addition to setting $S = V$.
$R$ must be large enough to cover the cost $C$ less the gains on the toehold
$(V-P_0) t_b q$, i.e., $R = C - (V-P_0) t_b q$. Then $A = (V - P_0) q - C$,
which is equal to the total social surplus. Clearly the best possible bid
will require choosing $(V^*, C^*)$, the business plan that maximizes total
social surplus subject to being feasible in the face of the market liquidity
constraint. That constraint allocates at least the proportion $1-t_m$ of the
total added token value to the $T_1$ token holders, who are free riders because they
bear none of the cost of adding value. The constraint may bind in the
optimization that determines $(V^*,C^*)$. Choice of $t_m < 1$ therefore may
preclude execution of the optimal business plan from a social perspective.\\

\textbf{Corollary 2.} \emph{If the market liquidity constraint, $t_f + t_b
\leq t_m$, is binding with $t_m <1$, then the bidder will choose a business
plan that falls short of creating the greatest possible social surplus.}\\

Suppose there are $n$ bidders and that bidder $i$'s best bid is $A_i^*$. In
an English auction, the highest bidder will prevail at the second highest
bidder's submitted auction parameter. The highest bidder can prevail with a
bid of at most $A_1 = A_2^*$, and typically, $A_1 < A_1^*$, the highest
bidder's best possible bid. Then, subject to the market liquidity
constraint, the highest bidder can increase $R$ from its level $R^* = C^* -
\Pi_{t_b}$, where $\Pi_{t_b} = (V^* - P_0) t_b q$ is the \emph{bidder's
toehold profit}, to $R^* + \min\{A_1^* - A_2^*, N^* - \Pi_{t_b} - C^* - F$\}.
The term $N^* - \Pi_{t_b} - C^* - F$ is the maximum possible value of $R$
that is feasible under the market constraint, given $F$, the minimum surplus
that must be delivered to the free riders.

This algebraic exposition can be visualized through a series of figures.

\begin{figure} [h!]
\captionsetup{labelfont=bf}
\centerline{\includegraphics [scale = .7] {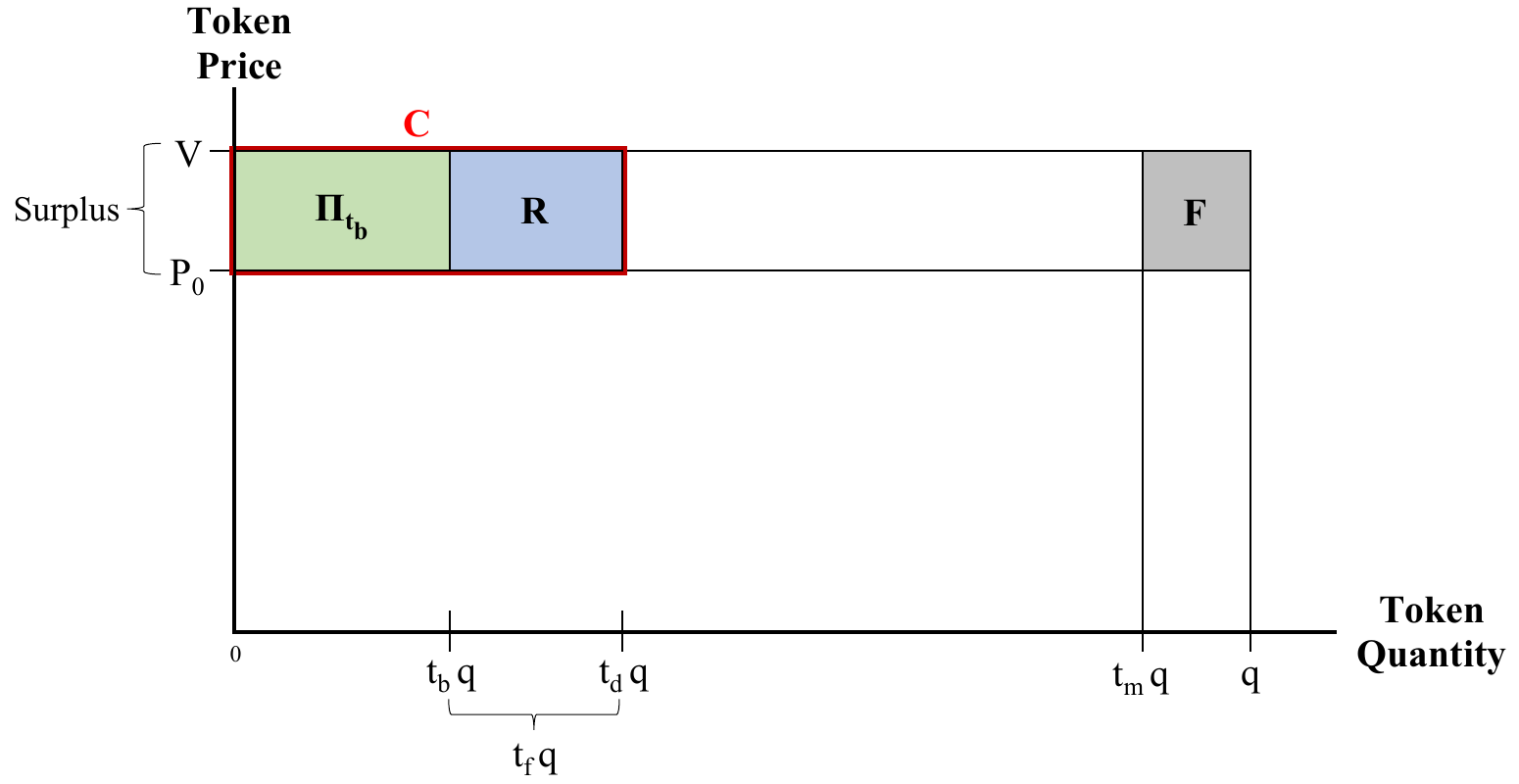}}
\caption{}
\end{figure}

\noindent In Figure 1, the business plan $(V,C)$ induces a value claim $S =
V$ and a surplus claim value of $R$ just big enough to cover the cost $C$,
which results in the winning bidder purchasing $t_f q$ tokens at $P_0$ from
the $T_1$ token holders through the freeze-out mechanism. This purchase
results in the bidder realizing added token value equal to the blue $R$ rectangle in
the figure. The green rectangle $\Pi_{t_b}$ is the added token value that the bidder
realizes from the toehold position $t_b q$. The bidder's total added token value is the
sum of the green and blue rectangles, which is equal in this instance to the
red rectangle representing the cost $C$. The area $F$ in the figure
represents the minimum added token value that must be granted to the other post-$T_1$
token holders, the proportion $1 - t_m$ of the total, and these free-riding
token holders also realize added token value equal to the unlabeled white rectangle between
the $R$ rectangle and the $F$ rectangle.

If $A_1^* > A_2^*$, the winning bidder will be able to appropriate part or
all of the added token value represented by the white rectangle. Figure 2 illustrates
the case where the winning bidder is able to appropriate part but not all of
it.

\begin{figure} [h!]
\captionsetup{labelfont=bf}
\centerline{\includegraphics [scale = .7] {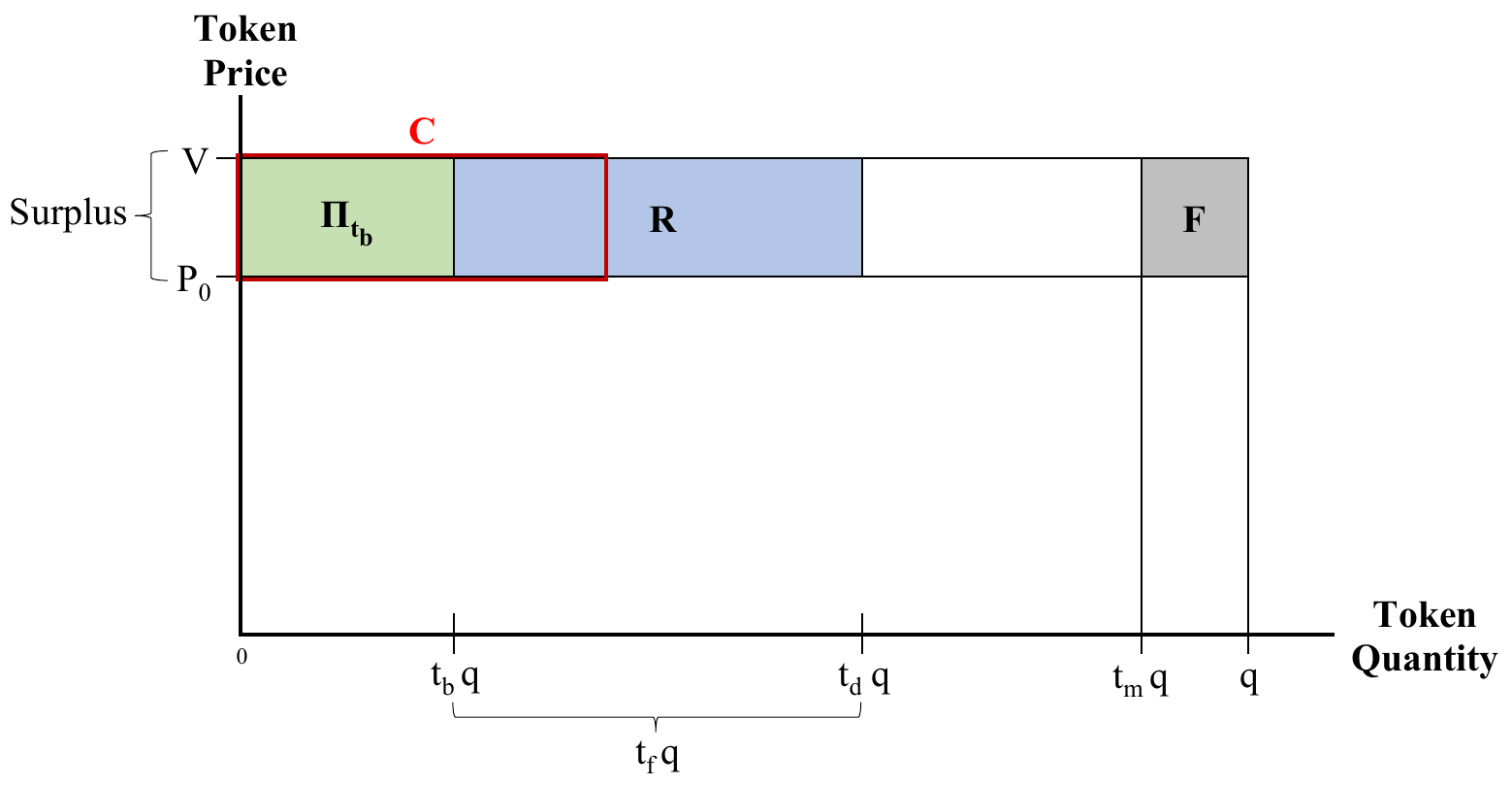}}
\caption{}
\end{figure}

These figures illustrate a key feature that enables the mechanism to work.
\citet{Burkart2015} show that in deterministic tender offer games the ability
to relinquish private benefits may be necessary for the existence of a
separating equilibrium. The auction mechanism here allows bidders to
relinquish benefits by choosing $R$, the surplus claim. A lower value of $R$
results in a stronger bid but a smaller amount of added token value accruing to the bidder. Because of the freeze-out
feature of the mechanism, the possibility that free-riding will undermine
signaling is eliminated. The bidder can specify an exact claim,
$R$, to added token value that otherwise might be inaccessible due to free riding. Combined with
the basic auction features that make revealing $V^*$ truthfully a dominant
strategy, the flexibility with respect to $R$ sets up an effective signaling
environment that leads to a separating equilibrium.

$A^*$ proxies for bidder types, and a bidder with a lower $A^*$ cannot
profitably mimic one with a higher $A^*$. For bidder $i$, the strongest
possible bid, $A_i^*$, is associated with zero profits. A higher bid results
in net losses. As the English auction unfolds, if each bidder $i$ moves up
until the level $A_i^*$ and then drops out, $V_i^*$ and $C_i^*$ will be
evident because the bid also reveals $t_b$.\footnote{As discussed in subsection \ref{efficiency}, it is possible that the second highest bidder submitting a bid with $A_2>A_2^*$ by understating the surplus claim required to break even is a profitable strategy. Although the bidder risks an operational loss if the bidder wins the auction, the bidder will retain a larger proportion of its toehold and earn the ensuing profit per token, $V-P_0$, on that larger position if it loses. The winning bidder faces a higher floor, $A_2 > A_2^*$, instead of $A_2^*$ and cannot claim as much of the available social surplus in that instance.} For the winning bidder, only
$V_1^*$ and an upper bound on $C_1^*$ will be evident because it is possible
that the winning bid, $A_1 < A_1^*$ so that $\Pi_{t_b} + R_1 > C_1^*$. In
that case, the winning bidder is receiving more added token value than is necessary to
induce that bidder to implement the best possible business plan. This
situation appears to fall short of the first best which requires that all
of the social surplus remains with the free riding token holders in order to
optimize initial and on-going investment in the DAO. But, as discussed in
subsection \ref{inv_impact}, the full picture is more complex.

\subsubsection{Bidding Intensity, Group Bidding, and Collusion}

Two elements that are relevant to the investment impact of the mechanism are
bidding intensity and the possibility of collusion among bidders.
\citet[Chapter 11]{Krishna2010} lays out basic points about collusion in
second price auctions. If there are $n$ bidders, then conditional on winning
the auction, bidder 1 expects to capture added value equal to $A_1^* -
E(A_2^* ~|~ n - 1 ~ \mbox{and}~ A_2^* \leq A_1^*)$. The expectation in this
expression is the expected value of the highest order statistic of the other
$n-1$ values of $A_i^* \leq A_1^*$. If a group of bidders collude, only the
bidder with the highest value among them will bid. That strategy eliminates
all the other bidders in the group from potentially lowering the highest
bidder's added value from winning by submitting the second highest bid. If the
collusion group is a subset of size $n_c \leq n -1$, then the expected
added value capture is with respect to a highest order statistic of a smaller
number of other bidders, $n - n_c - 1$ instead of $n-1$, and therefore is
higher. In the extreme situation in which all bidders collude effectively,
the highest bidder can prevail with a bid just above $P_0$, thereby
extracting the proportion $t_m$ of the total gain, the maximum possible
subject to the market liquidity constraint. The winning bidder then splits
the added value from collusion with the other colluding bidders, a division
that motivates them to be part of the collusion group.

Collusion, especially involving more than a few potential bidders, requires
costly coordination, but if it occurs, it will impact investment negatively.
Collusion results in winning bidders extracting more added value, reducing the
attractiveness ex ante for initial and on-going investment. There are some
situations in which collusion is a very plausible threat. For example, if
there is a strong outside candidate to add value through an innovative
business plan, the most likely competing bidders may be a handful of
identifiable ``insiders" who previously had been the most active in directing
the DAO. Small numbers make collusion easier to coordinate, and if the small
group includes the likely highest set of bidders, collusion is likely to be
very effective. The fact that DAOs have been characterized by low
participation rates and a few key players running the show suggests that this
situation may be common. Outside candidate collusion with the insiders
creates a substantial opportunity for reducing the social surplus that otherwise would accrue
to the other token holders.

A key parameter is the expected bidding intensity. If there is a continuum of
potential bidders who have values of $A^*$ that are dense in the interval
$[0, A_1^*]$, then it will be impossible to collude and the winning bidder
will only be able to cover cost. Investment incentives will be maximized. If
there are only a handful of potential bidders, then collusion will be a
bigger danger, and, even in the absence of collusion, the winning bidder will
be likely to walk away with considerable extra social surplus to the detriment of
investment.

Bidding intensity and the viability of the mechanism itself depends on
\emph{capital market adequacy}, the depth and development of capital markets.
Making a bid through the mechanism requires deposits of the same order of
magnitude as the total pre-bid value of the DAO. For established DAOs such as
Uniswap, Maker, or Compound, deposit amounts of a billion dollars or more may
be required.\footnote{As of November 24, 2024, these three DAO governance
tokens had market capitalizations of approximately \$6.4 billion, \$1.5
billion, and \$580 million respectively. \citet{Coinmarketcap_2024}.} A deep
bench of institutional investors such as venture capital firms, private
equity firms, and hedge funds is required to make the auction mechanism
robust and create bidding intensity. At least on the venture capital front, a
large number of such firms already exist and are very active with respect to
cryptocurrency projects including many
DAOs.\footnote{\citet{Crowdcreate_2023}.} These firms would be a ready source
of funding for auction bidders. The temporary contestable control created by
the auction mechanism, if implemented, might itself spawn specialist
institutions similar to hedge funds in the current corporate landscape that
combine portfolio investment with selective activism in
governance.\footnote{\citet{Yin_2023} provide an empirical analysis. Such
hedge funds typically engage in activism with respect to only a small portion
of their portfolios. \citet{Burkart2021} summarize the evidence that takeover
activism aimed at or resulting in changes in corporate control generate
higher target and activist returns.}

One aspect of the auction mechanism is highly relevant to the possibility of
collusion. There is nothing that prevents group bidding. A pool of outsiders,
of insiders, or a mixture of both might combine into a bidding entity. It is
possible to implement this combination via a smart contract, with the parties
contributing the required stablecoins to fund bidding and then jointly
holding the control position if their bid prevails. Group bidding may
facilitate collusion, but it also may add valuable bidders who otherwise
would not participate. For example, an entrepreneur with promising innovative
ideas for running the DAO but with limited resources could form a bidding
entity with private equity or venture investors who could fund the project.

As described in \citet{Krishna2010}, collusion in English auctions typically
is illegal or subject to civil penalties in the non-digital world, but is it
hard to see how collusion might be detected and policed in a code feasible
manner. Group bidding via smart contracts may be detectible but whether it is
merely collusive or has bid-formation advantages or is a mixture of the two
would be a complex inquiry, requiring something akin to legal processes.
Although there are mediation and adjudication tools available apart from the
legal system, using these tools adds an additional element of complexity and
requires trust in the tools.\footnote{\label{disputes} A prominent example of
a mediation and adjudication tool is Kleros. \cite{KlerosH_2023}.
\cite{Greig_2022} describes how Kleros operates by rewarding jurors for their
performance based on a Schelling point criterion. Kleros is active, with
multiple open cases. \cite{KlerosB_2023}. \citet{Guillaume_2022a} discuss use
of dispute resolution vehicles, including Kleros, for DAOs generally, noting
that they are most effective when remedies can be executed on-chain because
the parties have assets at risk there. In the case of collusion, some
relevant parties may not.}

\subsubsection{Toehold Reporting and Post-auction
Market Dangers} \label{toehold_danger}

The basic auction mechanism requires that the bidder report any toehold
position as part of the bid for two reasons. First, an accurate
report of the toehold is necessary for the mechanism to work as intended. The mechanism is designed to rank bidders based on how much social surplus they can offer the other token holders. To calculate the social surplus offered by a particular bid, it is necessary to subtract two quantities from the total added token value, $(S-P_0)q$, that the bidder asserts will be created. One is $(1-t_b)(S-P_0)q$, the portion of the added token value that the bidder will capture through the bidder's toehold position. The second is the bidder's surplus claim. These two quantities flow to the bidder and not to the other token holders. For a rational bidder, the sum of the two quantities will at least cover the bidder's costs, and to the extent the sum exceeds the two quantities, the bidder will appropriate some of the social surplus. If the bidder hides the toehold position by not reporting it, the bidder will be claiming to transfer more social surplus to the other token holders than is actually the case. Part of the claimed transfer of social surplus to the other token holders actually will be retained by the bidder. The bid will have an artificial advantage. If this bidder wins, it may be the case that some other bid promised to deliver more actual social surplus to the other token holders, subverting the operation of the auction mechanism.\footnote{All this is evident from Figures 1 and 2 in subsection \ref{optimal_bid} supra. In the figures, the white portion plus the grey portion labelled F of the top rectangle is the amount of social surplus the bidder is claiming to transfer to the other token holders. The green portion of the top rectangle is the portion of the total added token value that the bidder will realize on the bidder's toehold position. Hiding this green portion by not reporting it represents falsely that it is social surplus being transferred to the other token holders, making it appear to be part of the white portion of the upper rectangle. The bid appears to be better than it really is.}

Second, the true value of the toehold proportion, $t_b$, is necessary to enforce the market liquidity condition $t_b + t_f \leq
t_m$, which guarantees that at least $(1-t_m) q$ tokens are held by others,
creating a pool of market liquidity. Relevant to both purposes, nothing in
the bidding mechanism itself creates an incentive for a true report, and, in
fact, the incentive is to cheat by hiding some of the toehold to artificially
inflate the bid, $A$.

Effectively addressing potential toehold concealment depends both on the
\emph{identifiability} of the control party's token holdings, defined as the
ability to associate positions to the control party, and on the costs of
doing so. Complicating matters, any identification method must be EV-robust.
The potential threat is from hidden ownership, which is, as discussed in the
introduction, the obverse of empty voting. Instead of voting with no economic
ownership, hidden ownership involves economic ownership without voting. It
can easily be implemented via an equity swap between the control party and a
token holder. In exchange for the economic interest in the DAO, including
token appreciation, the control party offers another economic position such
as treasury bond returns to the token holder. The token holder continues to
own the token and will be identified as the owner on chain, but, secretly,
the control party is able to overstate the social surplus being offered to the other token holders by not reporting the associated economic position as part of the control party's toehold. This maneuver may be very hard to detect. In effect, a much deeper level of identification is required, one that reaches beneath formal ownership.

Potential concealment of part or all of a toehold position is not the only
threat that creates a need for identification. There are possible dangers
arising from post-auction market manipulation by control parties. If the
market is thin enough, the control party can engage in focused buying to
drive the market up to levels that substantially reduce the value or surety
deposit obligations at the time when those obligations are measured and enforced, and then engage in selling to reverse the temporary
position, similar to a pump and dump.\footnote{The value deposit and surety deposit obligations are measured and forfeiture of part or all of the deposits is possible at control transition points. Subsections \ref{post_auction} and \ref{subsequent_auctions} describe and discuss the various transition points including the end of control periods by time limit,  subsequent auctions that end control periods before the time limit is reached, abandonment of control during a control period, and the success termination of a control period based on attaining the token target price established by the value claim for a sustained period of time. \label{manipulation}} Whether the market is robust or not
depends both on the market capitalization of the DAO token and on the
proportion of tokens not held by the control
party.\footnote{\citet{Hamrick2018} provide substantial evidence that pump
and dump manipulations are much harder, as measured by the induced percentage
price increase, for heavily traded, high market capitalization
cryptocurrencies.} If the market liquidity constraint is binding, the control
party will exit the auction with the maximum proportion of $1 - t_m$ of the
total tokens outstanding, a situation that increases the danger of
manipulation because of reduced market liquidity stemming from a lower
traded-token supply. If the manipulation danger is salient enough, then for
the auction mechanism to work it may be necessary to bar the control party
from the post-auction market. Enforcing this bar would rely on an
identification procedure that could operate effectively during the control
period in the face of control by the control party.\footnote{Identification in the face of market manipulation not only would facilitate countermeasures internal to the DAO but also would empower external actors who have prosecution and enforcement authority. In the United States, for instance, the Commodities Futures Trading Commission has the power to police market manipulation for almost all publicly-traded cryptocurrencies. See subsection \ref{regulation} infra.}

For an identification method to work in a decentralized framework it must be
code feasible as well as effective. We consider some possible methods in what
follows, concluding with what appears to be the most promising one. We focus
on the problem of concealment of a toehold position. The discussion applies
in an obvious way to attempting to enforce a bar on market participation.

One method of identification is a \emph{bounty system}. Any party that
discovers and proves that the winning bidder concealed part of the toehold
rather than reporting it would be awarded a number of tokens with current
value equal to the value of the concealed part of the toehold at $T_0$, the
time of concealment, while simultaneously burning the corresponding quantity
of concealed tokens held by the control party. If the value of the concealed
tokens at the time the bounty is granted is less than their $T_0$ value, the
shortfall can be made up by burning part of the control party's token deposit
or by creating debits against the control party's stablecoin deposits.

Could a bounty system be effective and code feasible? DAO positions appear as
a set of public addresses and token quantities. Commercially available
technologies to trace and attribute token ownership exist and have a
significant degree of effectiveness at what may be feasible
cost.\footnote{See \citet{YaffeBellany2023} (discussing Chainalysis).} But
reliance on outside commercial parties inhibits code feasibility. The DAO may
have to implement the bounty system via contracts with outside teams. Doing
so effectively while the control party has control of the DAO may not be
possible. The task is complicated by the possible need to resolve disputes
about the veracity of identifications claimed by the bounty
hunters.\footnote{Dispute resolution is more difficult when the parties do
not have accessible on-chain assets at risk. See note \ref{disputes} supra.
An approach such as requiring good faith deposits from bounty hunters might
be required to implement a dispute resolution mechanism that is code
feasible.} And possible hidden ownership presents a major challenge to the
effectiveness of bounty systems.\footnote{Bounty hunters do have some
possible strategies in the face of hidden ownership. Counterparties to the
hidden ownership position used for concealment have an incentive to
collect the bounty by disclosing the failure to report, potentially earning 
additional tokens at the expense of the control party without violating the
underlying contract. Bounty hunters may angle for a cut by advertising this
opportunity broadly and offering assistance.} Although a bounty system may be
too difficult to implement, it is potentially a very powerful deterrent.
Successful bounty recovery is a disaster for the winning bidder, involving loss not only of potential added token value, the motivation for concealment, but also of the base value, $P_0$.

One set of approaches to address control party position reporting issues
centers around using know-your-customer \emph{registration systems} that make
token holders identifiable. Requiring registration of all token positions at
all times would be very costly both initially and in the face of continual
revisions as tokens are traded between parties. Registration might also raise
privacy considerations requiring costly zero-knowledge proof or other
technologies to make identities private yet verifiable. Nonetheless, use of
registration limited to subsets of token holders and to particular points in
time might be useful and cost-effective as we describe in what follows.

One particularly promising way to address the toehold reporting problem is to
use a \emph{flush sale} variant of the mechanism. This variant is
characterized by the following differences from the mechanism described so
far:
\begin{enumerate} [label=\arabic*)]
\item \emph{Revised purchase deposit}. The purchase deposit is $D_p^f =
    (1-t_b) P_0$, covering all the tokens other than the reported toehold position consisting of $t_b q$ tokens.
\item \emph{Flush sale}. The DAO Code uses this deposit to purchase all
    of the tokens other than the reported toehold at price $P_0$. This purchase
    implements the flush sale.
\item \emph{Adjust the token deposit}. The DAO Code adds or subtracts
    tokens from the token deposit to adjust that deposit by the quantity
    $t_f q$.
\item \emph{Token auction}. At the end of the basic auction, the DAO Code
    initiates a token auction, selling $(1 - t_d) q$ tokens using a
    hard-coded auction technology that aims at revenue maximization.
    Purchases by the control party are barred, enforced to the extent
    feasible by a registration system.
\item \emph{Registration}. Parties can register through a know-your-customer process to demonstrate that they are not the control party
    or related to the control party.\footnote{Registration can be made
    consistent with privacy through approaches such as requiring a
    zero-knowledge proof confirmation that they are not a restricted
    party linked to the winning bidder. For example,
    \cite{Rosenberg_2022} describe a zero-knowledge proof credentialing
    system that they call ``zk-creds." In one instantiation, parties
    embed their passport in a zero-knowledge privacy layer that allows
    proof of identity.} All registered parties are eligible to
    participate in the token auction. The set of $T_1$ token holders who
    register will be eligible to receive surplus from the auction.
\item \emph{Treatment of token auction surplus or deficit}. A token
    auction deficit caused by an average token auction price below $P_0$
    remains a liability of the DAO. Any token auction surplus is
    distributed pro rata at a specified \emph{flush sale surplus
    distribution date} to the set of registered $T_1$ token holders based
    on their relative $T_1$ holdings. This date is set by the DAO Code to
    give sufficient time for $T_1$ token holders to register before the
    flush sale surplus distribution date if they did not do so prior to
    the token auction.
\end{enumerate}
The flush sale variant attempts to address the problem of non-reporting of
the full extent of the winning bidder's token position. Some or all of it is
hidden among the $1- t_b$ proportion remaining after the proportion $t_b$ is
declared and deposited. The flush sale is just that: It flushes out any
hidden control party positions among that remaining proportion by forcing
sale of the entire remaining proportion at $P_0$.

It is likely that the auction purchases will be at prices significantly
higher than $P_0$. For that reason, even without the success of a bounty
system, registration, or similar measures, the winning bidder is faced with a
potential reduction in added token value on any positions that were concealed even if
accompanied by corresponding repurchases after completion of the token
auction. In contrast, declaring and depositing the pre-auction positions puts
them in a safe harbor that allows the winning bidder to collect 100\% of any
added token value that arises. The registration system is crucial for establishing
these incentives because the system ensures that any added token value from the auction
flows to $T_1$ token holders other than the control party. Registration for
participation in the token auction also plays a role because market trading
after the auction is likely to be accompanied by substantial additional
demand for tokens and a corresponding sharp price increase similar to what
happens in a successful initial public offering in equity markets.
Registration requirements block control parties from enjoying these gains.
Use of a registration system aimed at blocking control parties from making
auction purchases is likely to be particularly effective compared to
alternatives such as relying on bounty hunters, and, because it would be limited in scope, relatively low cost.

Flush sales implemented using registration can be combined with bounty hunting or other measures aimed both at the pre-sale concealed positions and at any purchases during the token auction. More than one such measure may be employed simultaneously. For the flush sale variant to be successful, registration and any other accompanying measures combined with the safe harbor aspect of reporting must make concealment in order to improve the bid, $A$, unprofitable, unattractive, or infeasible. The stakes are significant. To the extent that the winning bidder can evade reporting the toehold position, it will gain an artificial bidding advan\-tage equal to the amount of added token value associated with the concealed position. Auction efficiency is imperiled because a bidder with a larger toehold may prevail over another bidder who has a superior business plan.

\subsubsection{The Role of the Toehold and Activism Strategies}
\label{toehold_role}

Toeholds play a crucial role in the current market for corporate control.
\citet{Burkart2021} show that under current law both activists and tender
offerors profit primarily from toeholds in the face of Jensen-Meckling and
Grossman-Hart free riders respectively. Furthermore, \citet{Burkart2015}
prove that the toehold can play a signaling role because choosing the level
of the toehold is a way to claim or relinquish control benefits.

Under the basic auction mechanism, there is no reason for a potential bidder
to accumulate or add to a toehold if the potential bidder is confident of
winning the auction and toehold reporting is enforceable. The size of the
toehold does not affect bidding strength or potential profits. When a
potential bidder is contemplating initiating an auction, buying more tokens
to accumulate a bigger pre-auction toehold is a dominated strategy. Doing so
will only drive up the price during accumulation when the potential bidder
can use the freeze-out feature of the basic auction to force sale of the
tokens at the lower pre-accumulation price. Furthermore, any potential
signaling role of the toehold is extraneous because the basic auction allows
the bidder to claim or relinquish control benefits directly.

One remaining question revolves around the role of traditional activism in
which the activist buys a toehold and then engages in a costly campaign to
influence management. In the case of a DAO, the targets of influence would be
the control party during a control period and the most active governance
parties otherwise. If the activist has a concrete value-increasing plan that
the activist could implement alone or after assembling a bidding group, the
activist is better off initiating an auction if the activist is confident of
winning. The activist could acquire shares more cheaply through the auction
freeze-out feature than by buying a toehold and would not have to incur
campaign costs to convince the control party or the active governance parties
to adopt the activist's business plan.

Any potential bidder, activist or not, has to consider the possibility of
losing the auction that they initiate. Losing the auction means there was a
higher bid, and if the winning bidder's business plan is credible, then a
losing bidder can profit from a toehold position. However, there are better
ways than a toehold position to insure against losing the auction. Suppose
that the potential bidder contemplates a bid that includes a value claim
$V_b^*$ when the token price is $P_0$. Simultaneously with making a bid, the
potential bidder can purchase a suitable quantity of out-of-the-money call
options with a strike price of $V_b^*$. If the bidder loses and the winning
bid includes a value claim $V^* > V_b^*$, then the call option will yield
what would have been the toehold profit. If the bidder wins, the bidder can
liquidate the call option position, most likely at a profit because it is
plausible that the token price will be greater than $P_0$ when the auction
closes at $T_1$, and possibly substantially greater.

\subsubsection{Auction Efficiency and Market Liquidity}\label{efficiency}

Like second price auction frameworks in other contexts, the English auction that embodies the basic mechanism here has strong efficiency properties. The bidder who can produce the most social surplus wins, with appropriation of that surplus by the winning bidder limited by the level of the second highest bid. As discussed, there is another limitation: The mechanism is subject to the market liquidity constraint. This constraint limits the added token value available to the winning bidder, reducing it by $F=(1-t_m)(V^*-P_0)q$, the minimum amount that must accrue to free riders. Consider the project $(V^*, C^*)$ that creates the highest possible social surplus $\psi^*=N^*-C^*$ and produces total added token value $N^*=(V^*-P_0)q$. If $\psi^* = N^* - C^* > 0 > N^* - C^* - F$, then this best possible project will not be implemented because it cannot cover its cost and also distribute $F$ to the free riders.

The purpose of the constraint is to enable the token market to operate continuously, which is important because the market value of the DAO tokens is a key input for the auction mechanism. Other approaches are possible, but it is not clear that they would resolve the trade-off between having a continuous active market and avoiding the danger that some socially valuable projects will be precluded by free riding that limits the available added token value. One possibility is a variant of the flush sale and subsequent token auction discussed in subsection \ref{toehold_danger}. If the market liquidity constraint is binding, then the flush sale at price $P_0$ consists of the $(1-t_m)q$ tokens that are not the held by the winning bidder at the conclusion of the auction. The variation applied when the constraint is binding is that instead of surplus from the token auction being directed to the $T_1$ token holders, it would be paid, up to the amount $F$, to the winning bidder, giving that bidder some or all of the added token value that previously would have flowed to the free-riding $T_1$ token holders. At the same time, the token auction would reestablish market liquidity.\footnote{A bounty system or alternative enforcement mechanism would be required to block the winning bidder from participating in the token auction directly.} It is unclear what portion of the missing added token value, F, that the winning bidder would in fact realize from this forced sale plus token auction approach. Market participants might be skeptical of the ability of the winning bidder to fulfill the $V^*$ value claim by bringing the token value to that level. Addressing this situation by allowing the winning bidder to delay the forced auction to create time to demonstrate the value of the business plan would create a period with no market prices.

There is another possible source of inefficiency, which we will call ``toehold overbidding." \citet{Burkart1995} shows that in an English auction, it may
be optimal for participants with toeholds to bid higher than their
valuations. Overbids create a danger of overpaying if the overbidder wins,
but they also may push up the winning auction price, benefiting the toehold
position if the overbidder loses. \citet{Burkart1995} proves that, in the context of his model, for sufficiently small overbids, the potential toehold benefit outweighs the danger of loss from winning the auction with an overbid.\footnote{The potential gain on the toehold is of the order of magnitude of the overbid while the potential loss from the overbid is of the order of magnitude of the overbid squared. \citet{Burkart1995} assumes a private values setting in his main inquiry. \citet{Bulow1999} show that toehold overbidding tends to be a much more serious problem in common values situations.} 

The mechanism developed here eliminates the incentive to make toehold overbids. Bidders are competing to offer as much added token value to the other token holders as possible. There two ways to offer more added token value: through a higher value claim or a lower surplus claim. However, as long as the value deposit forfeit function satisfies the inequality condition in Lemma 1, bidders are better off increasing their bids by lowering their  surplus claims. But doing so has no impact on the corresponding value claims and therefore cannot cause competing bidders to earn more on each share of their toehold positions by overbidding.\footnote{The result in Lemma 1, that making excessive value claims is a dominated strategy, remains intact.}

Nonetheless, there is a strategy analogous to toehold overbidding that can have similar effects under the mechanism: \emph{surplus claim underbidding},  lowering the surplus claim below the point that causes the best business plan of the bidder to break even. The goal is to cause the winning bidder to also lower its surplus claim. That smaller surplus claim reduces the proportion
of $T_1$ token holdings that must be sold to the winning bidder at $P_0$ and
increases the corresponding proportion on which the $T_1$ token holders,  including the bidder who engaged in surplus claim underbidding, can
earn the full amount of added token value.

Successful surplus claim underbidding is socially desirable because more social surplus is shifted to the $T_1$ token holders, which results in a positive investment impact while not affecting the result that the best business plan is put into effect. Unsuccessful surplus claim underbidding,
however, has negative social properties. The winning bidder implements an
inferior business plan. To the extent of the underbid, additional apparent
social surplus flows to the $T_1$ token holders based on the value deposit forfeit function, which will result in a positive investment impact. However, this impact is larger than is optimal, because at least part of the surplus is apparent rather than real, being in excess of what would be generated even by the best business plan.\footnote{Investment is over-encouraged if we look at this auction event in isolation. It is possible that the transfer in excess of actual social surplus generated by this type of event offsets the failure to shift the full amount of social surplus to the $T_1$ token holders in other auctions. In short, the transfer in excess of social surplus may be second-best optimal in some instances.} It is not clear how these two impacts balance out across auctions, but it is clear that a net negative effect is possible.

\subsection{Post-auction Operation} \label{post_auction}

In this subsection we examine post-auction operation absent a subsequent
auction, deferring discussion of subsequent auctions to the next subsection.
After the basic auction, the winning bidder is now the \emph{control party}
because the empty votes held in the dynamic vote pool plus the votes
associated with the token deposit give the winning bidder more than a
majority of the DAO votes outstanding. Control persists for a \emph{control
period}, the time period between the end of the basic auction and the time
when the DAO Code closes the dynamic vote pool. In the meantime, the control
party's token deposit, value deposit, and surety deposit are locked in
appropriate smart contracts, to be released in part or entirely when certain
conditions are met.

There are various circumstances under which the DAO Code will end the control
period arising from a previous basic auction. One such circumstance,
discussed in the next subsection, is when there is a later, supervening
auction. However, there are situations in which the DAO Code will end the
control period in the absence of a supervening auction. We examine one such
situation next.

A \emph{Success Termination} occurs if the token price reaches at least $S$ on a sustained basis. ``Sustained basis" must be defined in a manner that is code feasible. For example the criterion might require that the token price as determined by some group of price oracles maintained by the DAO averages at least $S$ for 30 days and sustains a value at or above $S$ consecutively for at least 10 of those days. A Success Termination has the following consequences:

\begin{enumerate} [label=\arabic*)]
\item The dynamic vote pool is closed, removing all the empty votes
    associated with the pool from the control party, and ending the
    control period.
\item The entire value deposit and the entire surety deposit are
    returned to the control party.
\item The token deposit is released from the applicable smart contract
    and returned to the control party.
\end{enumerate}

After a Success Termination, the DAO returns to the default governance state with the former control party holding at least $t_d q$ tokens, where $q$ is the quantity of tokens outstanding when the previous auction ended at time $T_1$. If the former control party wants to retain control and holds less than 50\% of the outstanding tokens at termination, that party will need to initiate and win a new basic auction. Otherwise, an \emph{open period} begins, during which the control aspects of the auction mechanism do not operate.

\subsection{Subsequent Auctions} \label{subsequent_auctions}

Subsequent auctions to an initial basic auction are of two types. First, any
party can initiate a basic auction either prior to the termination of control
from a previous auction or at a time which is not within a control period.
There is nothing new about a basic auction initiated outside of a control
period, but when a basic auction is initiated during a control period, the
DAO Code will need to specify how that auction interacts with the control
framework in place following the previous auction. Second, the DAO Code
specifies potential \emph{periodic auctions}. These potential auctions
commence when the \emph{control period} reaches the time limit specified by
the DAO Code.

\subsubsection{Basic Auctions During a Control Period} \label{A_during_CP}

During a control period, the entire control structure including a token
deposit, a value deposit, and a dynamic vote pool along with all of the
associated parameters will be embodied in one or more smart contracts. If a
party initiates a basic auction during a control period, then there are two
scenarios: (i) the control party bids in the auction and wins; (ii) some other
party wins the auction whether or not the control party bids.
In the first scenario, there will be an \emph{Auction Reset} implemented via a set of \emph{Auction Reset Steps}. These same Auction Reset Steps apply to periodic auctions when the previous control party bids and wins. Suppose that the new auction begins at $T_{c\,0}$ and ends at $T_{c\,1}$ and that the parameters from the previous auction were $S$, $P_0$, $D_v$, $D_p$, $D_s$, and $t_d q$. Denote the corresponding parameters emerging from the new auction and the associated winning bid with a superscript ``$w$," except that, as stated above, we assume for convenience that the number of tokens outstanding remains at $q$. In this first scenario as well as subsequent ones, we will use a reference price $P_{ref}$ equal to the token price as of the time $T_{c\,0}$ when the auction begins, which we denote $P(T_{c\,0})$.\footnote{Use of a price at a single point in time to settle value and surety deposits or to set auction parameters creates the danger that various parties will attempt to manipulate market prices at that point of time, perhaps combined with choosing the timing of actions such as initiating an auction or abandoning control. See supra note \ref{manipulation} and accompanying text in subsection \ref{toehold_danger}. There are at least two levels of response. First, in some jurisdictions such as the United States, market manipulation is illegal and is policed by regulators. Second, the mechanism itself can choose a reference price that estimates the price on the crucial day but is harder to manipulate, such as using an average of prices during a time interval surrounding that day. There are many possible approaches, and avoiding manipulation may have costs in terms of accuracy. In addition, in some cases, if capital markets are developed enough and active, market forces may constrain manipulation. For example, engaged and informed short sellers may defeat attempts to manipulate prices upward. It is clear that the best mechanism responses to the manipulation danger may depend on capital market quality and the features, including market liquidity, of particular DAOs. We leave exploration on these fronts to future work.} 

We now develop a \emph{transitional forfeit function} that determines how much of the value deposit the control party forfeits when the control period ends with another auction. Because there are multiple instances in which this situation can arise, it is convenient to specify the function and explain its operation in advance to avoid tedious repetition. 

Define $\Delta \phi((D_v,P_0,S,S^w,P(T_{c\,0}))$, the \emph{value deposit forfeit differential}, which we will use for the case $P(T_{c\,0}) \ge P_0$: 
$$\Delta \phi = \Big( \phi(D_v,P_0,S,P(T_{c\,0})) - \phi(D_v,P_0,S^w,P(T_{c\,0}))  \Big)_+$$
where $(Z)_+ = \max (0,Z)$ is the positive part of the quantity or function $Z$. The purpose of this differential is to reduce the value deposit forfeit amount imposed at the end of the initial control period to the extent that the control party has recommitted to honoring the surplus claim, $S$, from the first auction.\footnote{Note that in most cases, the forfeit amount $\phi(D_v,P_0,S,X)$ will monotonically decrease in $S$. A lower value claim will result in a lower forfeit if the same token value $X < S$ is realized by executing the business plan. Thus, the difference that defines $\Delta \phi$ will be positive if and only if $S^w < S$. In that case, the control party has not fully reaffirmed the commitment to reach $S$, which was embodied in the value deposit $D_v$ required in the first auction. Thus, some loss of that previous value deposit is appropriate. \label{higher_value_claim}} The first term is the forfeit amount that would be imposed at the end of the initial control period if there were no adjustment. The second term is an appropriate credit in terms of the initial control period value deposit forfeit function based on the fact that the control party is committing to attain at least the token value $S^w > P(T_{c\,0})$ during the new control period. The commitment is credible because failure to close the gap between $P(T_{c\,0})$ and $S^w$ during the new control period will result in a corresponding loss of part or all of the value deposit applicable to that period. 

Define, $\phi^*((D_v,P_0,S,S^w,P(T_{c\,0}))$, the \emph{transitional forfeit function}, as follows:
\begin{equation} \nonumber
\phi^* =
\begin{cases}
                \Delta \phi((D_v,P_0,S,S^w,P(T_{c\,0})) & \text{if } P(T_{c\,0}) \ge P_0 \\                
                D_v  & \text{if } P_0 > S^w > P(T_{c\,0})\\  
                \Delta \phi(D_v,P_0,S,S^w,P_0) & \text{if } S^w \ge P_0 > P(T_{c\,0}).
        \end{cases}
 \end{equation} We have already described the operation of this function in the first case, $P(T_{c\,0}) \ge P_0$. In the second and third cases, $P(T_{c\,0}) < P_0$. The fact that the token price outcome was less than $P_0$ raises the possibility of value destruction by the control party. However, the two cases are quite different on this dimension, and, as a consequence, quite different forfeit amounts are appropriate. In the second case, the outcome below $P_0$ is not accompanied by any commitment by the control party to bring the token value to or above $P_0$ during the new control period. Thus, it is appropriate for that party to suffer full loss of the previous value deposit, including any baseline loss penalty, $\phi_0 = D_v - (1-t_d)(S-P_0)q$, which, as discussed in subsection \ref{optimal_bid}, is useful for deterring value destruction.

In the third case, by making a value claim $S^w \ge P_0$, the control party is committing to raise the token price to at least $P_0$ during the new control period, subject to losing part or all of the new value deposit if the commitment is not fulfilled. The costly commitment to restore the lost value suggests that the control party was not motivated by value destruction during the initial period. As a result, imposing the baseline loss penalty is not appropriate. Instead, the starting point is $\phi(D_v,P_0,S,P_0)$, a forfeit amount equal to the value deposit, $D_v$, less the baseline loss penalty. We reduce this forfeit amount by crediting $\phi(D_v,P_0,S^w,P_0)$ based on reasoning similar to the first case: The control party has committed to reaching $S^w > P_0$, making a penalty appropriate only with respect to the gap $[S^w,S]$, a gap encompassing the portion of the failed commitment from the initial period that has not been renewed in the new period.\footnote{The appropriate level of credit is $\phi(D_v,P_0,S^w,P_0)$, not $\phi(D_v,P_0,S^w,P(T_{c\,0}))$, a larger amount that includes the new value deposit commitment in the extra range $[P(T_{c\,0}),P_0]$ given that $P(T_{c\,0}) < P_0$. This extra range was not part of the control party's original commitment in the first auction, a commitment that only involved values greater than or equal to $P_0$. The goal is to credit appropriately for settling up that original commitment.} After subtracting the credit, we arrive at exactly the forfeit amount $\Delta \phi(D_v,P_0,S,S^w,P_0)$ stated above for the third case. 
 
We have explained the transitional forfeit function based on the first scenario for a subsequent auction, the one in which the previous control party wins the auction. The same transitional forfeit function applies in the second scenario in which the previous control party loses the subsequent auction. Much of the reasoning for applying this particular function in the second scenario is the same or analogous, but we leave the details to later. With the transitional forfeit function in hand we are ready to state the Auction Reset Steps, the transitional rules that apply in the first scenario.

The Auction Reset Steps are: 
\begin{enumerate}[label=\arabic*)]
\item \emph{The purchase deposit}. The control party made a purchase
    deposit of $D_p^{w} = \max \left\{ \frac{R^{w}}{S^{w}-P_0^{w}}
    P_0^{w}, 0 \right\}$ as part the current auction. The appropriate
    smart contract deploys the purchase deposit to buy $t_f^{w} q$ tokens
    at $P_0^{w}$ from the $T_{c\,1}^{w}$ token holders pro rata based on
    their $T_{c\,1}^{w}$ token holdings. The DAO Code adds the purchased
    tokens to the control party's token deposit.
\item \emph{The token deposit and dynamic vote pool transition}. The
    token deposit shifts from $t_d q$ to $t_d^{w} q$ after appropriate
    token deposits or withdrawals by the control party. The token deposit
    is retained in the appropriate smart contract during the subsequent
    control period. The dynamic vote pool continues in place, adjusting
    the parameter $t_d$ to the new value $t_d^{w}$.
\item \emph{The value deposit transition}. When the new auction concludes
    at time $T_{c\,1}$, two control party value deposits are outstanding: a
    \emph{previous value deposit} $D_v$ arising from the prior auction, and
    a \emph{current value deposit} $D_v^w$ that the control party
    made at the time of submitting the control party's final bid in the
    just completed auction. The current value deposit is retained. Out of
    the previous value deposit, the amount forfeited by the control party is $\phi^*((D_v,P_0,S,S^w,P(T_{c\,0}))$, the applicable value of the transitional forfeit function. This forfeited amount is paid to the $T_1$ token holders from the previous
    auction pro rata based on their $T_1$ token holdings. The remainder is returned to the control party.
\item \emph{Treatment of the current surety deposit}. Parallel to the
    value deposit, there are two surety deposits outstanding at the end
    of the auction: a \emph{current surety deposit}, $D_s^{w}$, and the
    \emph{previous surety deposit}, $D_s$. The current surety deposit is
    retained in an appropriate smart contract.
\item \emph{Treatment of the previous surety deposit}. Define the
    following three shortfall parameters:
    \begin{enumerate} [label=(\roman*)]
    \item The \emph{value shortfall}: $H = \max\{ (P_0 - P_0^w) q, 0
        \}$.
    \item The \emph{adjusted value shortfall}: $H^* = \max\{H -
        D_v, 0 \}$.
    \item The \emph{bid shortfall}: $B = \max\{(P_0 - S^w) q, 0 \}$.
    \end{enumerate}
    The DAO Code returns $\max\{D_s - B, D_s - H^*, 0\}$ to the control
    party and transmits the rest to the $T_1$ token holders from the
    previous auction pro rata based on their $T_1$ token holdings.
\end{enumerate}

As will be discussed more extensively in subsection \ref{value destruction}
below, the surety deposit is designed to protect against value destruction by
bidders who gain control of the DAO. We make some observations for the scenario
we are considering in which the control party wins the new auction. Loss of
part or all of the surety deposit only comes into play if: (i) the value of
the DAO has fallen below $P_0 q$ at the time the new auction commences,
creating a value shortfall $(P_0 - P_0^w) q$; and (ii) this value shortfall
is greater than the previous value deposit $D_v$. Even if these two necessary conditions are met, if $S^w \geq P_0$,
the entire previous surety deposit is returned. The control party can
therefore block losing any part of the previous surety deposit by bidding at
least $P_0$. However, in the case where there is a value shortfall $(P_0 -
P_0^w) q > 0$, the entire value shortfall becomes part of the current value
deposit and is at risk. The potential of loss remains to that extent.

In the second scenario, the control party loses the new auction. The winning
bidder is treated according the Basic Auction rules with respect to all
aspects, including deposits, the creation of a dynamic vote pool, and the
initiation of a new control period. Again denoting the parameters emerging
from the new auction and the associated winning bid with a superscript
``$w$", the losing control party is treated according to the following
\emph{Control Party Auction Loss Steps}, which include exactly the same
treatment of the previous value and surety deposits as under the Auction
Reset Steps, the applicable transition rules for the first scenario:

\begin{enumerate} [label=\arabic*)]
\item \emph{Closing the previous dynamic vote pool}. The previous dynamic
    vote pool is closed, removing all the empty votes associated with the
    pool from the losing control party, and ending the control period
    associated with that party.
\item \emph{Return of the previous token deposit}. The previously
    existing token deposit is released from the applicable smart contract
    and returned to the losing control party.
\item \emph{Treatment of the previous value deposit}. The losing control party forfeits $\phi^*((D_v,P_0,S,S^w,P(T_{c\,0}))$ of the previous value deposit, the amount specified by the transitional forfeit function. The forfeited amount is paid to the $T_1$ token holders from the previous
    auction pro rata based on their $T_1$ token holdings. The remainder is returned to the losing control party.
\item \emph{Treatment of the previous surety deposit}. Define the
    following three shortfall parameters:
    \begin{enumerate} [label=(\roman*)]
    \item The \emph{value shortfall}: $H = \max\{ (P_0 - P_0^w) q, 0
        \}$.
    \item The \emph{adjusted value shortfall}: $H^* = \max\{H -
        D_v, 0 \}$.
    \item The \emph{bid shortfall}: $B = \max\{(P_0 - S^w) q, 0 \}$.
    \end{enumerate}
    The DAO Code returns $\max\{D_s - B, D_s - H^*, 0\}$ to the control
    party and transmits the rest to the $T_1$ token holders from the
    previous auction pro rata based on their $T_1$ token holdings.
\end{enumerate}

Several features of these rules are noteworthy with respect to their role in creating participant incentives through the auction mechanism. First, in the normal situation in which $\phi(D_v,P_0,S,X)$ is monotonically decreasing in $X \in [P_0,S]$, the losing control party is rewarded by a lower potential value deposit forfeit amount to the extent the outcome $P(T_{c\,0})$ that the party delivers as of the beginning of the subsequent auction is higher. Second, if the winning bid in the subsequent auction includes a value claim of $S$ or higher, then the losing control party is off the hook.\footnote{See note \ref{higher_value_claim} supra.} Similarly, if the losing control party's bid includes a value claim of at least $S$, in effect renewing a commitment for the DAO token to reach this level, then unless opposing bidders can win with a bid below $S$ because they have significantly lower costs, the control party will not lose any of the value deposit.

As noted, the treatment of the previous surety deposit is exactly the same as
in the scenario in which the control party wins the auction and the Auction Reset
Steps apply. We add some observations with respect to losing control parties,
leaving a comprehensive discussion of the surety deposit for later in
subsection \ref{value destruction}. A winning bid with a sufficiently high
value claim will guarantee return of the previous surety deposit, the
previous value deposit, or both. In particular, $S^w \geq P_0$ results in a
return of the previous surety deposit, and $S^w \geq S$ guarantees the return
of the entire previous value deposit. The return of these
deposits under the applicable conditions is justified by the following
reasoning. First, the condition $S^w \geq P_0$ based on a value claim by an
independent bidder suggests that the control party did not engage in
significant value destruction, justifying return of the entire previous
surety deposit. Second, losing the auction means the control party loses the 
opportunity to bring the token price to $S$ or above and thus redeem the
previous value claim. At the same time, given $S^w \geq S$, the winning
bidder is asserting that it will be able to do so. And the losing control
party's value-building efforts or identification of an opportunity may have
been part of the basis for the winning bidder's value claim of at least $S$.
Under these circumstances, withholding return of any portion of the previous
value claim is not consistent with creating appropriate incentives for
bidders who are capable of adding value to the DAO through their efforts. Similar reasoning justifies return of part of the previous value deposit as specified by the transitional forfeit function when $S^w < S$ but $S^w \ge P_0$.

Finally, note that the portions of the value deposit and surety deposit that
are not returned to the losing control party remit to the $T_1$ token holders
from the previous auction regardless of whether or not they have retained
their $T_1$ token holdings. Potential return of the deposits attaches to the
\emph{holders} and not to the \emph{tokens}. \label{signal} As a result, potential return of
the deposits does not affect the market value of the tokens. Otherwise,
buyers would price in the probability of deposit returns, increasing the
value of the tokens above the level that would follow from future business
prospects alone. This kind of distortion would be problematic because the
auction mechanism relies on the token price being a meaningful signal of the
value of the DAO.

The control party may want to relinquish control in the absence of a
supervening auction by \emph{Abandonment}. In this scenario, the Control Party
Auction Loss Steps apply, setting $S^w = P_0^w$, as if the control party lost
the auction to a bid that included a value claim equal to current market
value at the time of Abandonment.

\subsubsection{Periodic Auctions}

If the entire control period elapses without early termination due to success
or a supervening auction, then the DAO Code triggers a new basic auction. The
new auction starts at the end of the control period, and the parameter $P_0$
is set at the value of the DAO governance token, $P_0^p$, as of that time.

If there is at least one bidder, then the approaches of the previous subsection
apply directly. If the control party wins, the DAO Code implements the
Auction Reset Steps. If the control party loses or does not bid, then the
Control Party Auction Loss Steps apply. If no one bids, then the control
party is treated as losing an auction in which the winning bid included a
value claim of $P_0^p$: The Control Party Auction Loss Steps apply, taking
$S^w = P_0^p$ and $P_0^w = P_0^p$ in terms of the notation used in the Steps.

It is worth considering some situations in which a periodic auction is
triggered by the passage of time. Suppose that the control party is still
confident of attaining a sustained token price of $V^*$ as the business plan
plays out. The fact that a periodic auction is taking place means that to
date only a lower sustained level has been attained. In this situation, the
control party has a strong incentive to bid in the periodic auction with the
bid including a renewal of the previous value claim so that $S^p = S = V^*$
where $S^p$ is the new value claim applicable to this particular periodic
auction. This bid means that the control party will not lose any part of the previous value or surety deposits. That will also be the case if another
bidder wins with an even higher value claim.

A second situation is where the control party engaged in extreme value
destruction, dropping the initial token value from $P_0$ to some small but
nonzero value. If no one including the control party bids, then in the standard case, the control party
will lose $P_0 q$. The control party can forestall loss of both the value
deposit and the surety deposit by submitting a new bid that renews the
previous value claim.\footnote{The value deposit from the previous auction is
returned to the control party because the new value claim is at least as high
as the value claim from the previous auction. The surety deposit from the
previous auction is returned to the control party because the bid shortfall
is zero. But the value deposit for the new auction exceeds the sum of these
previous two deposits.} Given that the DAO has nominal value, it is a near
certainty that there will be no competing bids. But unless the control party
puts resources and effort into the DAO, the low value will persist until the
next periodic auction. In effect, the value and surety deposits will be stuck in the DAO
and effectively lost.\footnote{The DAO Code could create a maximum number of
sole-bid control periods to close out the stuck position, but this might
entangle honest control parties with projects that take a long time to prove
out.} A more profitable strategy for the control party is to allow its
control to expire whether or not there is a bidder in the periodic auction,
thereby releasing its entire remaining token deposit, which the control party
can sell to at least recover some money. In fact, the control party may want
to abandon the control position before even reaching a periodic auction.
Abandonment has the same consequences because the same Control Party Auction
Loss Steps apply that pertain to loss of a periodic auction in the case where
there are no bidders in that auction.

\section{Further Evaluation of the Mechanism} \label{more}

\subsection{Impact on Investment} \label{inv_impact}

As discussed in section \ref{roots}, an important consideration is the impact of
the sequential auction mechanism on initial investment in DAO projects. The
first best is to offer winning bidders only the amount of added token value required to
cover their costs and execute their business plans. If initial investors
expect that they will retain more social surplus in subsequent auctions, they will
pay more for the venture in the first place. A mechanism with the highest
expected retained social surplus will result in the highest possible funding for the
DAO at the start-up point as well as the highest possible on-going investment
value.

One question is whether full surplus extraction (``FSE") is possible. At
present, it appears that FSE is attainable only under prescribed, relatively
narrow conditions. In particular, \citet{Cremer1988} show that if types are a
finite set, the types are correlated, the joint probability distribution of
types is known to both the mechanism designer and the bidders, and that
probability distribution meets certain conditions, then it is possible to
design an FSE mechanism. A large literature surrounds and extends this
result, encompassing, for instance, a continuum of types, applicable here
because $V^*$ and $C^*$ are real
numbers.\footnote{\citet[pp.~120-128]{Borgers2015} contains a good general
discussion.}

These possibility results are almost certainly not relevant here. The
assumption that the joint probability distribution of types is known to both
the mechanism designer and the bidders is highly implausible. \citet{Fu2021}
show that sampling without prior knowledge from this distribution may suffice
to achieve FSE, but is unclear how that sampling could occur in the context
here.

Robert Wilson's statement of the so-called ``Wilson doctrine'' in
\citet{Wilson1987} is particularly apt in the context of designing mechanisms
for DAOs:
\begin{quote}
Game theory has a great advantage in explicitly analyzing the consequences of
trading rules that presumably are really common knowledge; it is deficient to
the extent it assumes other features to be common knowledge, such as one
agent's probability assessment about another's preferences or information. I
foresee the progress of game theory as depending on successive reductions in
the base of common knowledge required to conduct useful analyses of practical
problems. Only by repeated weakening of common knowledge assumptions will the
theory approximate reality.
\end{quote}
Trading rules are code feasible, and reducing them to code makes them common
knowledge among all potential bidders. But it is unreasonable to expect other
common knowledge elements to be present and available to the mechanism,
especially because essential elements such as the bidding participants and
the surrounding circumstances may not be known until shortly before or during
the auction itself.

The problem of extracting surplus is equivalent to designing a revenue
maximizing auction. Here the ``revenue'' is the added token value that remains from the
business plan after subtracting the amount of added token value, $R + \Pi_{t_b}$, that accrues to
the winning bidder. We have used an English auction, which is a second price auction because the winning bidder pays the amount of the second highest bid. Various other auction types may
raise more revenue under particular circumstances. For example, if bidders
are risk averse, a first price auction will raise more revenue but may also
result in an inefficient outcome, which in our case means that the business
plan creating the most social surplus may not be selected. Furthermore,
adding some additional realistic assumptions such as bidders whose outcome
possibilities differ because of different levels of skill, makes it unclear
whether a first price auction indeed would raise more revenue than an English
auction.

Another consideration is that sealed-bid first price auctions may avoid the
collusion problems that can arise in second price auctions. Collusion
coalition parties in a sealed-bid first price auction can defect by bidding
above an agreed upon low price chosen to enable the highest valuing bidder to
prevail at that price and thus collect the largest possible amount of added token value
for the coalition. In a second price auction, the highest valuing bidder will
submit their best bid, and other coalition members cannot win the auction by
deviating. Deviating only reduces the joint added token value for the coalition. The
potential considerations in choosing among auction forms are myriad, and we
do not attempt a detailed discussion here.\footnote{A good starting point is
\citet{Krishna2010}.} Instead, we simply rest on a claim that given current
knowledge, the English auction is a strong candidate with respect to
efficiency and raising revenue compared to other standard
auctions.\footnote{\label{common_values} We have been assuming a ``private
values" setting with asymmetric information in which bidder business plans
are known only to each bidder and bidder valuations of the plans are
independent. See note \ref{private_values} supra. If, instead, there is a
common values element in which bidders have different signals concerning the
value of the current operation and bid based on perceived undervaluation at
price $P_0$, then the efficiency and surplus-distribution of various auction
methods are potentially different than under the private values assumption.
However, the English auction performs relatively well in a common values
setting. Under certain assumptions such as the average crossing property, the
English auction remains efficient in the sense that there is an ex-post
equilibrium in which the bidder who can add the most value will win.
\cite[pp. 134-143]{Krishna2010}. In this setting, the English auction tends
to raise more revenue than other approaches, including sealed-bid first price
and second price auctions. \cite[pp. 97-100]{Krishna2010}. In the context
here, raising more revenue equates to extracting more added token value from the
winning bidder, one of our desiderata.}

An English auction has other strong traits worth mentioning. An English
auction is ``obviously strategy proof'' in the sense of \citet{Li_2017}. In
intuitive terms, it is an obvious dominant strategy to bid until one's true
value is reached and then drop out, at least when there is a continuum of
bidder types.\footnote{As discussed in subsection \ref{efficiency}, when
there is not such a continuum and especially when there is a small number of
bidders, there may be an incentive to overbid.} This quality minimizes the
possibility that the auction will go awry due to the inability of some subset
of bidders to understand what they should do under the mechanism.\footnote{On
a similar note, \citet{Neeman_2003} demonstrates the robustness of the
English auction to the seller's degree of ``Bayesian sophistication'' with
respect to setting a reserve price in particular private-values
environments.}

Another feature that relates to surplus extraction is the reserve price. The
mechanism requires $S \geq P_0$, setting a reserve price equal to the token
price at $T_0$, the start of the auction period. This reserve price makes
market dynamics and market efficiency important. Token holders anticipate
that the DAO project might be improved in the future, and an efficient token
market implies capitalization of these expectations in the form of a higher
token price. To the extent that the expectations are cogent, the auction will
involve much less social surplus. In effect, some social surplus was already extracted and
accrued to the investors through the impact of the capitalized expectations.
If the expectations err to the high side, the token will be overvalued,
potentially blocking implementation of even the highest social surplus business
plan. However, when no innovations emerge after a period of time, it is
likely that the token price will fall until an innovation becomes feasible
under the auction mechanism. It is not clear how prevalent and potent these
phenomena might be, but it is clear that they might have big impact on
surplus extraction and thus on initial investment.\footnote{One possibility
is to create a multiple-stage model, such as the one in
\citet{OrdonezCalafi2022}, with the initial investments in the DAO occurring
at the first stage and auctions at later stages. This type of model would
permit examining the impact of the degree of surplus extraction in the
auctions on initial investment and how that impact depends on various
parameters. We leave development of such a model or analogous ones to future
work.}

\subsection{Preventing Value Destruction} \label{value destruction}

The danger of value destruction arises from the possibility that the control
party will take the equivalent of a negative economic position in the DAO,
one that would increase in value if the DAO flounders, and then will use
control to cause the DAO to perform poorly or even to fail entirely. The
value of the control party's token deposit consisting of $t_d q$ tokens would
decline in the event of value destruction, but the deterrent effect of this
deposit is not EV-robust. The control party can remove the economic aspect of
this position, converting the position to empty votes, through a variety of
empty voting strategies. As pointed out in subsection \ref{empty}, some of these
strategies are not costly and would be hard to detect by any available means,
much less by code feasible methods. The rest of the control party's votes are
from the dynamic vote pool, and these votes are empty votes by
design.\footnote{The creation of empty votes through a dynamic vote pool is
extremely useful. It allows the winning bidder to secure control with less
than a majority of the tokens, thereby enabling the $T_1$ token holders to
secure a greater proportion of the social surplus generated by the bidder, which
will have a positive investment impact. } Thus, the control party can arrange
to secure control entirely through empty votes.

Both the value deposit and surety deposit are at risk if the control party
engages in value destruction. Although the value deposit also plays a role in
eliciting bids and incentivizing post-auction performance, the sole purpose
of the surety deposit is to supplement the value deposit in order to deter
value destruction. Recall that the surety deposit is:

$$T_s =
\max \left\{ P_0 q (1-\gamma) - D_v, 0 \right\}$$

\noindent where $\gamma$ is a choice parameter. $D_v$ is the value
deposit. If $\gamma = 0$, then the surety deposit is set such that the sum of
the surety deposit and the value deposit is $P_0 q$, the entire value of the
DAO at the beginning of the auction. If the control party engages in value
destruction that drives the value of the DAO to $0$, then the control party
will forfeit both deposits in their entirety. Loss of these deposits is
EV-robust. As discussed in subsection \ref{EV_robust}, the control party
would have to pay substantial sums to construct a derivative position that
covered loss of both deposits.

The surety deposit is useful in deterring potential value destruction, but it
also may deter bidders who intend to create value. Despite being confident
that their business plans will move the DAO value up from $P_0 q$ to $S q$,
there may be an interlude in which the DAO governance token price falls below
$P_0$ before the business plan proves itself. An auction during that
interlude, including a periodic auction if the control period ends during the
interlude, creates the danger of loss of part or all of the surety deposit
under the applicable Auction Reset Rules or Control Party Auction Loss Rules.
In the deterministic setting we have been assuming, there will be no problem
if the control party remains confident in the business plan. In that case the
control party will reinstate the previous bid, which includes a value claim
$S > P_0$, ensuring the return of the entire surety deposit. On the other hand, if the control
party no longer has enough confidence in the business plan to bid with a
value claim of at least $P_0$, a bid that would trigger return of the entire
surety deposit, then there is a loss. But this loss is appropriate because it is due to the failure of the business plan, a failure that drove the expected token price below the original starting value, $P_0$.\footnote{In the assumed deterministic setting, only auction-initiated business plans and the timing and degree of their success determine token prices. Low token price outcomes may occur due to stochastic elements not associated with the cogency of the business plan. As discussed in Appendix
A.2, an appropriate response to that possibility is particular ex-ante adjustments to each of the deposits as well as to the value deposit forfeit function.}

By choosing the parameter $\gamma \leq 1$, it is possible to calibrate the
surety deposit appropriately to the level of threat posed from potential
value de\-struc\-tion by a control party. In particular, $\gamma$ can be set
to a value consistent with the operating and financial environment of the
DAO. We have seen that if $\gamma = 0$ is chosen, then between the surety
deposit and the value deposit, a control party engaging in value destruction
risks forfeiting up to $P_0 q$, the entire pre-auction value of the DAO,
through loss of the surety and value deposits. Circumstances may dictate that
$\gamma$ does not need to be that low. For example, if value destruction is time-consuming, and there are many potential, informed bidders available to step in, then it is likely that value destruction will be prevented, at a loss to the control party, by a new auction.

The design of the value deposit and its interrelationship with the surety deposit are important considerations with respect to value destruction. Because value destruction necessarily involves driving the token value below the baseline price, $P_0$, a party intent on value destruction will forfeit the entire value deposit for certain. A larger value deposit is a stronger deterrent against value destruction. 

Increasing the value deposit may have undesirable collateral consequences. In particular, as discussed more fully in subsection \ref{EV_robust}, there is a danger of creating excessive and socially counterproductive post-auction incentives for control parties to expend costly effort aimed at increasing the value of the DAO. It is here that the baseline loss penalty described in subsection \ref{basic} is particularly valuable. That penalty is a fixed amount that is levied if and only if the token price outcome is less than $P_0$. As discussed in that subsection and subsection \ref{EV_robust}, this penalty can accompany the standard value deposit forfeit function, which allows the designer the freedom to mold the forfeit amounts for outcomes greater than or equal to $P_0$ in any way desired to achieve the best possible set of post-auction performance incentives. The baseline loss penalty will deter value destruction of any magnitude, no matter how small, and increasing it does not affect the post-auction performance incentives of parties not intent on value destruction. 

Finally, note that if deterrence fails and value destruction is successful, forfeiture of the deposits fully compensates the token holders other than the control party. They are in at least as good of a position as they would be if the claimed business plan had been executed successfully.

\subsection{Undervaluation and Treasury Raid Scenarios}

Potential ``treasury raid" scenarios provide an interesting perspective on
how the mechanism would operate. DAOs typically have treasuries consisting of
cryptocurrency tokens that can be used for further development of the DAO.
Treasuries can store retained earnings as well as funds received from
investors. Treasuries may be under the direct control of token holders who
can direct their use by voting on proposals.\footnote{In some cases, there
are other parties who have that formal power or a veto on token holder
proposals for spending treasury resources. In one common organizational
structure, the DAO is embodied as a Swiss Foundation. The Foundation acts a
wrapper, creating legal personhood for the DAO, with concomitant limited
liability. Under Swiss law, the Foundation must have formal legal control over
the DAO treasury. Although the Foundation may in practice typically defer to
token holder votes, there is the power to veto vote outcomes if they involve
disbursing treasury funds.} Direct control by token holders creates the
danger of ``treasury raids." The most pernicious example of a treasury raid
is one in which a party gains control of the DAO and then transfers the
entire treasury to itself by voting positively on a suitable proposal. The
party is stealing the other token holders' stakes, and the term ``raid" along
with its negative connotations is apt.

Other cases and situations are not so clear. A relevant parameter is whether
the token market capitalization is above or below the value of the treasury.
The ``underwater" situation in which the market capitalization is
substantially below the value of the treasury is consistent with multiple,
contradictory possible realities, each of which calls for a different
perspective and optimal response. We consider three possible realities that
illustrate challenges for the mechanism in addition to the case of a direct
raid on the treasury. These three realities and the direct raid possibility
can occur whether or not token prices are underwater, but starting with the
underwater case makes the challenges clearer.

First, there may be an expectation that the parties currently guiding the
DAO, although good-intentioned, will engage in foolish business plans,
gradually wasting the treasury assets. The foolish business plans may be the
result of the exhaustion of the original guiding purposes of the DAO. In this
case, the social optimum is to end the DAO and distribute the treasury to the
token holders, permitting re-investment of the treasury assets in more
promising projects. In a situation of capital market adequacy, there will be
``vulture investors," who facilitate this result via a takeover or through
activism in the face of reluctant guiding or control
parties.\footnote{\label{Rook_DAO} Vulture investors already exist in the
cryptocurrency space. The saga of Rook DAO, recounted by \citet{Gilbert_2023}
and \citet{Nelson_2023}, is an example. Rook DAO encountered an internal
conflict and was trading below the value of its treasury. Co-founders  and
members of the core team proposed shifting 75\% of the tokens to themselves
to continue the project, leaving 25\% to the rest of the token community.
Activist investors facilitated an increase in the community's share to 60\%.
\citet{Liu_Arca_2023} details the claimed role of Arca, one of the activist
investors. As described in \citet{Arca_2023} and \citet{Dorman_2020}, Arca's
operations include some that resemble a hedge fund that both makes portfolio
investments and engages in activism.}

Second, the guiding parties might have a very promising business plan but are
keeping it secret from competitors. Current operations that lay the
groundwork for that business plan may appear substandard to market
participants, resulting in a market capitalization less than the value of the
treasury. In this case the socially optimal outcome is for the guiding
parties to continue in control until the business plan can play out and
reveal its superiority.

Third, the guiding parties may not be totally good-intentioned, and may be
diverting treasury assets to themselves short of an explicit distribution of
the entire treasury that could be characterized as a single ``raid" incident.
Some of the diversions may be ambiguous or subtle, such as overpaying DAO
contractors that are related to the guiding parties. Significant diversions
might cause the DAO to underperform noticeably, resulting in an underwater
token price. Diversions are tricky from a social optimality perspective. The
diverted assets are not lost, only redistributed. Furthermore, if the
diversions are anticipated at the time of investment, investors can give the
guiding parties a lower stake for the same amount of funding, nullifying the impact of the diversions on investor returns.

Assuming capital market adequacy, the mechanism can address all of these
situations effectively. Suppose first that the token value is underwater. Then a
bidder can initiate an auction at just above the current token value, $P_0$,
when the value per token of the treasury is $P_{treasury} > P_0$. If the
bidder wins, the bidder can terminate the DAO and distribute the treasury pro
rata to the token holders. Capital market adequacy ensures that the winning
bid will be close to $P_{treasury}$ and that most of the social surplus will flow to
the existing token holders.\footnote{The ``business plan" of liquidating and
distributing the treasury and hedging the treasury value over the required
short interval of time is unlikely to have high costs, and because the
situation is so clear, it is likely that there will be many potential bidders
and that most of the potential gain will be bid away, accruing to the existing token
holders instead. Bidder deposits will be effectively collateralized by the value of
the treasury.} Therefore, it is unlikely that the token value will be
significantly underwater, and the situation in which the guiding parties are
engaging in foolish projects will be terminated by liquidation of the DAO.

If, instead, the token value was underwater because the guiding parties have
a superior business plan that requires secrecy, capital market adequacy means
that these parties can obtain funding in confidence to outbid parties intent
on liquidation. The other token holders receive at least their pro rata share
of the treasury and possibly also some of the social surplus from the envisioned
superior business plan. If the guiding parties \emph{think} they have a
superior business plan but cannot convince any funding party of that opinion,
then upon liquidation, the guiding parties at least receive their share of
the treasury, possibly amounting to enough to implement their business plan
through a new project.\footnote{The outcome in the Rook DAO events described
in note \ref{Rook_DAO} supra appears to represent these kinds of splits
between parties intent on implementing a business plan and parties who want
to exit through a liquidation that brings the DAO value up to the level of
the treasury. The Rook DAO treasury assets were divided between two such
groups of parties. The bidding competition central to the mechanism creates a
way to divide the treasury assets more cogently, with the key question being whether
the parties who want to continue can outbid the liquidation value. If they
can, the other parties receive at least the pro rata treasury value of their
tokens and likely more. If they cannot, they receive their share of the
treasury and can pursue the project with those funds.}

The mechanism provides a corrective to the diversion situation. A bidder can
initiate an auction with the ``business plan" being carrying out the DAO
operations without the diversions. This business plan will increase the token
value of the DAO, creating room for the bidder to profit. At least some of
the added token value will flow to the other token holders.

Finally, the mechanism will tend to cause the aggregate token value to equal
or exceed the value of the treasury. As a result, a treasury raid executed by
using the mechanism to gain control will not be profitable. The DAO value
will fall by an amount that at least approximates the missing treasury
assets, and the winning bidder will lose that degree of its deposits, with
the lost deposits flowing to the other token holders, reversing the stealing
inherent in the treasury raid.

\subsection{Post-auction Incentives and EV-robustness} \label{EV_robust}

At the end of the auction, ignoring any additional token or token-derivative
positions taken on during the auction period, the winning bidder, now the
control party, holds $t_d q$ tokens on which that party hopes to earn added token value and
faces the task of avoiding loss of part or all of the value deposit by
executing the business plan in order to move the token price up to $V^*$, the
\emph{target price}. The control party's incentives to execute this
task depend on the value deposit forfeit function and the party's actual token
position. 

Under the standard token deposit forfeit function and assuming that holding $t_d q$ tokens is the control party's net position, a one dollar increase in the token price will result in a $t_dq$ dollar increase in the token holdings and a $(1-t_d)q$ drop in the value deposit obligation. The net result is that the control party is better off by $q$ dollars for every dollar increase in the DAO's token value, which is exactly the increase in the value of the DAO. In this case the control party has exactly the correct incentives to put effort into the project because the control party will capture 100\% of the gains and suffer 100\% of the losses from taking or refraining from effort. There is no Jensen-Meckling problem. The control party will behave as if it is a 100\% owner, comparing the costs of effort to 100\% of the potential gains.

Nearly costless hedging equivalent to a temporary sale of part or all of the toehold may be available through the use of derivatives or otherwise. In the case of a control party intent on the business plan and on increasing token value to the target price, any such hedging will have the undesirable consequence of reducing the amount of token gain that the control party will capture. For control parties intent on value destruction, the opposite is the case. Hedging avoids any loss on the toehold as a result of the value reduction, and giving away the upside in a hedge is costless for such a party.  A hedge equivalent to a sale sets up an empty voting position, exactly what the control party intent on value destruction desires.

It is clear that no arrangement is going to be EV-robust with respect to
token holdings. It is easy enough to hedge them in a way that is equivalent
to a sale. But the same cannot be said of the value deposit and security
deposit, which are key deterrents to value destruction and simultaneously
incentivize achievement of the business plan. These deposits are EV-robust.
Hedging the value deposit requires purchasing the opposite derivatives
position, which in the case of the standard value deposit forfeit function effectively means buying a put with a $V^*$ strike price and selling a
put with a $P_0$ strike price.\footnote{See note \ref{options note} supra.} This position would be expensive in the sense that it would cost a significant proportion of the deposit. With respect to outcomes below $P_0$, a large put position with appropriate strike prices would be required to offset potential forfeits involving both the value deposit and surety deposit. This put position also would be costly and possibly difficult to assemble without moving the token price significantly downward while assembling the position.

A major advantage of the mechanism is that control itself is completely EV-robust. Contesting
parties can only succeed by being the highest bidder in the auction enabled
by the mechanism. Holding or assembling empty votes has no impact. The
arbitrariness that can be associated with empty voting cannot occur.

There are factors that might induce a control party intent on successful implementation of a business plan to hedge part or all of its token position despite the loss of token gain that would flow from success. If the DAO position is a large part of the control party's portfolio, optimal portfolio management may require reducing the long position in the DAO. Any such reduction would mean that the control party is under-incentivized with respect to effort. Such a party bears 100\% of the cost of effort but reaps less than 100\% of any gains therefrom. If control parties hedge part or all of their token holdings in the DAO, there is a case for going beyond the standard token deposit forfeit function by making the value deposit forfeit amounts larger than $\phi(D_v,P_0,S,X) \approxeq (1-t_d)(S-X)q$ for each outcome $X$. For completely hedged control parties, full performance incentives are attainable by setting $\phi(D_v,P_0,S,X) = (S-X)q$. Then the control party is put in a position of a 100\% owner facing at least the full losses and gains for the DAO through the value deposit mechanism, which the control party cannot evade by hedging. However, if this token deposit forfeit function is imposed, then control parties who hedge less than their full token position in the DAO will be over-incentivized to perform, creating the danger that they will expend costs that exceed the corresponding increases in the value of the DAO.\footnote{For convenience of exposition, we have gone beyond the deterministic model to consider the impact of control party hedging at this point. Appendix A.2 discusses hedging strategies more generally and distinguishes between different types of control parties. If the control party is a large institutional investor or is funded by one, hedging the DAO token position may be less urgent with respect to portfolio balancing and may not be worth the potential loss of gains on the tokens. Choosing the form of the value deposit forfeit function and other aspects of the DAO may depend heavily on the nature of the particular DAO, including the characteristics of the likely control parties.}

\subsection{Code Feasibility}

The auction mechanism is code feasible by design. Deposits take the place of
derivatives that would require counterparties. The mechanism itself is an
easy-to-code algorithm.

Many DAOs contain structural governance limitations that specify which DAO
Code or DAO operation changes can be achieved by majority vote, and which
changes are subject to more stringent procedures. Control parties face the
same governance restrictions as other parties. Winning an auction only
guarantees that a control party can prevail with respect to any issue that is
subject to a majority vote ($>$ 50\%), even if the control party holds less
than half the tokens eligible to vote. If a higher voting percentage is
required, the control party can only succeed unilaterally through actual
token ownership, as is the case with all other parties.\footnote{Careful
consideration is required with respect to whether the control party can count
its empty votes from the dynamic voting pool as eligible token votes with respect to issues that require a supermajority vote to resolve. The answer to this question may depend heavily on which issues are involved and also upon the overall governance structure and philosophy of the particular DAO.}

There is a strong argument for making the code governing the auction rules
themselves immutable or nearly so.\footnote{If the auction provisions are
immutable, it will be impossible to improve the auction process itself, a
move that potentially would increase the value of the DAO. If some
flexibility is permitted, the danger of possible manipulation suggests
creating a substantial barrier to revision such as a very high supermajority
vote possibly combined with deposits or other vehicles that shift the risk of
any loss in DAO value upon the proposing parties.} As long as this code is
intact, auctions remain as a powerful tool to address governance issues and
to overturn any hold on the DAO by pernicious control parties, whether or not
the control originated from a previous auction. In addition, immutability of
the auction provisions prevents incumbents from distorting the auction
process to their advantage by making control shifts via the auction more
difficult or by creating rules that favor incumbent bidders.

The biggest difficulty regarding code feasibility is the task of bolstering
toehold reporting discussed in subsection \ref{toehold_danger} above. It is not
clear that detecting failures in toehold reporting is amenable to simple code
solutions. Possible measures include recourse to outside parties, such as
bounty hunters, or implementation of a suitable registration system requiring
know-your-customer verification. Perhaps the most powerful approach is to
build a flush sale into the mechanism that includes a registration system
that directs added token value from concealed control party positions away from the
control party. But all of these approaches require interactions with the
outside world and some level of trust that goes beyond secure code.

\subsection{Decentralization, Regulation, and Legal Aspects} \label{regulation}

A major question is the relationship of the mechanism presented here to
decentralization, a desideratum for DAOs. This question also has significant
legal and regulatory implications. The auction mechanism is a control device,
creating the possibility of temporary contestable control by a single party,
which itself may be an entity or group of persons. That may appear to be a
move toward centralization, but arguably it is the opposite. As mentioned,
many DAOs appear to be under the de facto control of a relatively small group
of active token holders, often connected to the founders, surrounded by a
mass of other token holders who rationally do not participate in governance.
Even if that control shifts, the shifts will be subject to the vagaries of
voting procedures and empty voting.

The mechanism improves on that situation by making control transitions more
effective, not the result of voting procedures that are subject to social
choice flaws and also not subject to the arbitrariness that can arise from
empty voting. Most important, control is \emph{continuously contestable}. Any
party can initiate an auction. Even a majority token holder cannot maintain
control in the face of an auction except by being the highest bidder. That is
a very different picture than a DAO with entrenched control held implicitly
or explicitly by founders or large token holders. It also is important to
keep in mind that the mechanism creates the possibility of alternation between
control periods and open periods in which the DAO reverts to a less organized
mode of operation, including voting approaches that may be valued for reasons
other than promoting operational efficiency.

Regulation of DAOs is not fully defined or developed and there often is
considerable uncertainty, especially in the United States where three major
quite different regulatory approaches are possible even for the near future.
Because these same three regulatory approaches are the major candidates
across jurisdictions, we consider the mechanism in light of each of them. We
present multiple examples of the approaches, but, to create a more enduring
discussion in light of the unsettled nature of cryptocurrency regulation, we
focus on the approaches in general rather than on specific current
instances.\footnote{The set of current regimes, even limited to the most
important ones, is likely to be obsolete within months, if not weeks, of
publishing this paper.}

After introducing and discussing the three regulatory approaches, we turn to the interaction of the mechanism with regulation. We begin with the most important and general point, that by providing additional protection for investors, adding the mechanism greatly strengthens the case for a much less burdensome regulatory regime which at the same time is likely to be more effective. Following that general discussion we consider ways in which the mechanism interacts with existing regulation under the three approaches.

\subsubsection{Three Regulatory Approaches}

A first approach is \emph{decentralization-focused} in the sense that the
degree of decentralization determines whether and to what extent DAOs and
other cryptocurrency applications are regulated. There are many variants that
fall into this category. Under the recently implemented European Union laws
governing cryptocurrencies, the key role of decentralization is explicit with
an exemption from regulation for the case in which ``crypto-asset services
... are provided in a fully decentralized manner without any
intermediary."\footnote{\citet{MiCA_2023} (Directive (22)).} In the United
States, the test set forth by the Supreme Court in \emph{SEC v W.J. Howey
Co.}\footnote{328 U.S. 293 (1946).} determines which cryptocurrencies will be
considered ``securities," subject to registration with the Securities and
Exchange Commission (``SEC") and on-going regulation by the Commission.
Cryptocurrencies that are not ``securities'' generally fall under the
regulatory ambit of the Commodities Futures Trading Commission (``CFTC") as
``commodities.''\footnote{A cryptocurrency can be both a ``security" and a ``commodity" under the applicable statures and regulations, creating joint regulatory oversight by the SEC and CFTC. Virtually all cryptocurrencies are considered to be "commodities," and the CFTC has regulatory authority to police fraud and market manipulation in spot markets for commodities. The SEC has the same authority with respect to cryptocurrencies that also are ``securities."}

One branch of the \emph{Howey} test requires investor dependence on the
``efforts of others" to make a profit as a necessary condition for a token to
be a security.\footnote{328 U.S. 293, 299 (1946).} If the token is decentralized, in the extreme
case just code running on its own, arguably there is no such dependence.
Furthermore, disclosure of the backgrounds and plans of the key managing
parties is a major objective of registration and on-going regulation under
the securities laws.\footnote{\citet{Hinman_2018}.} If there are no such
parties, regulation loses its primary rationale. In addition, if there are no
managing parties, it is unclear who would be available to comply with
regulation in any event.

An example in which the degree of decentralization affects the type of
regulation rather than creating a potential exemption from regulation is H.R.
4763, a recent bill passed by the U.S. House of Representatives, that would
create a comprehensive regulatory framework for
cryptocurrencies.\footnote{H.R. 4763 is known as the Financial Innovation and
Technology for the 21st Century Act (``FIT21"). H.R. 4763 was approved by the House
Financial Services Committee on July 26, 2023 and passed by a strong bipartisan majority of the full House on May 22, 2024, creating the possibility of enactment if approved by the Senate and signed by the President. \citet{HR4763}; \citet{HR4763pass}.} This legislation would
grant the CFTC primary jurisdiction over any blockchain network or
application that is functional and decentralized.\footnote{The proposal
defines both a ``decentralized network'' and a ``decentralized governance
system.'' In each case, the focus is on whether there is a person with
effective control. In the case of a ``decentralized network,'' there also is
the requirement that no party has 20\% or more of ownership or voting power.}
If the system is not yet functional or is not decentralized, then it is
regulated by the SEC. SEC regulation is much more onerous because it requires
formal registration of the project, along with periodic reporting
requirements. In contrast, unless the project is an exchange, CFTC regulation
is largely limited to addressing fraud and market manipulation. There are no
registration or reporting requirements. This difference is critical for DAOs
because DAOs lack centralized management, and, as a result, compliance with
registration and reporting requirements will be difficult if it is possible
at all.

A second kind of regulation can be described as a \emph{joint and several
responsibility approach}. Some party or set of parties \emph{must} be
responsible for meeting traditional regulatory requirements, and in the
absence of a designated party or parties, a wide variety of participants are
effectively at risk of being held responsible with ensuing penalties or
liabilities for the failure of the DAO to comply. There is a danger that this
approach would preclude decentralized operation entirely. In the United
States recent developments with respect to both the SEC and CFTC rules
governing exchanges are striking examples.

In a case involving Ooki DAO,\footnote{Commodity Futures Trading Commission
v. Ooki Dao (N.D. Cal., Dec. 20, 2022).} a commodities exchange operated
through a DAO, the CFTC claimed that Ooki DAO was an exchange that was
required to register under the Commodities Exchange Act and be subject to
regulation by the CFTC but that had failed to do so. The founders had
switched the exchange from operating as an LLC to a DAO, publicly stating
that the purpose was to avoid regulation. After settling with the founders,
the CFTC proceeded against the DAO itself despite the lack of identifiable
parties managing the DAO, with the result being a default judgment for money
damages in excess of \$600,000, an injunction, and an order requiring that
Ooki DAO end its internet presence.\footnote{\citet{CFTC_06_09_23}.} The CFTC
took the position that Ooki DAO was comprised of token holders who had voted
on any proposal, whether or not related to any regulatory matter. Because
Ooki DAO is an unincorporated association, these token holders most likely
will be jointly and severally liable for the full amount of the money
damages.

On the SEC side, developments are even more threatening. The SEC recently
reopened the comment period for proposed new rules covering the definition of
exchanges that are required to register with the SEC and comply on an
on-going basis with a complex set of regulations (the ``Reopening
Release").\footnote{\citet{SEC_Reopen}.} The Reopening Release was motivated
largely by the goal of clarifying the SEC's position on the applicability of
the rules to token-based decentralized exchanges, including those governed by
DAOs.\footnote{Id. at 6-7.} As part of the new rules, the SEC had proposed
expanding the definition of an exchange by adding a ``\emph{group of persons}
that constitutes, maintains, or provides a market place or facilities for
bringing together buyers and sellers of securities" to entities that can
comprise an exchange.\footnote{Id. at 6 (emphasis added).} Commentators
pointed out that for decentralized projects, code replaces intermediaries,
and a ``group of persons" could include a very wide range including
developers and miners. The SEC responded first with the general principle
that: ``The existence of smart contracts on a blockchain does not materialize
in the absence of human activity or a machine (or code) controlled or
deployed by humans."\footnote{Id. at 22.} The SEC then made clear the
astonishing breadth of parties that might be included in the ``group of
persons" collectively held responsible for securities law compliance:
interface providers, code developers, DAOs, validators, and issuers and
holders of governance tokens if they act in concert or share common
control.\footnote{Id. at 27.} Related more directly to DAOs, the SEC stated
that ``... significant holders of governance or other tokens ... could also
be considered part of the group of persons and thus an exchange if they can
control certain aspects of it."\footnote{Id. at 25.} The aspects included:
securities available for trading, requirements and conditions for
participation, determining who can share profits or revenues, or having the
ability to enter into legal and financial agreements on behalf of or in the
name of the market place or facilities.\footnote{Id. at 25.} The SEC was not
willing to exclude software developers, who, acting independently and
separately from the project, published code that later was picked up and used
in the project by an unrelated person. These parties were only ``less likely"
to be acting in concert.\footnote{Id. at 28.} The SEC also made clear that
liability exists for persons or entities that initially created or deployed
the system's code even though ``the system, once deployed, typically cannot
be significantly altered or controlled by any such persons."\footnote{Id. at
29.}

The SEC went on to discuss compliance costs. The SEC noted that ``factors
associated with certain technologies ... might increase compliance costs" and
that compliance might ''significantly reduce the extent to which the system
is `decentralized' or otherwise operates in a manner consistent with the
principles that the crypto asset industry commonly refer to as
`DeFi.'"\footnote{Id. at 122.}  The SEC singled out DAOs as a case for which
the highest possible compliance costs might occur.\footnote{The Commission
put it this way (Id. at 124-125): \begin{quote} The Commission preliminarily
believes that a reasonable case, in which the highest possible compliance
costs would result, would be a ... system that performs exchange activities
in part using smart contracts, but in which control over changes to the smart
contracts is given to a token-based voting mechanism, which may use
governance tokens as discussed above, and where the tokens are dispersed
among a large number of investors. \end{quote}} The SEC continued, stating
that ``the holders of governance tokens, or other tokens that carry voting
rights, may bear the responsibility of ensuring compliance with the system,"
suggesting that the token holders should create an organization or delegate
persons to undertake the activities required for compliance, putting money
into the DAO if the costs are not covered by fees.\footnote{Id.} Furthermore,
the token holders would have to have the power to alter the relevant smart
contracts by vote if necessary for compliance, and in the case of immutable
smart contracts, miners or validators would have to alter the smart contracts
via a hard fork.\footnote{Id. at 126-127.}

While it is true that these positions were asserted in a Release soliciting
additional comments, they are the responses to concerns raised by previous
comments suggesting that the scope of the Rules should be limited to create
an exemption or at least breathing space for decentralized projects,
including DAOs. And although the Rules only apply to exchanges, the same
reasoning would apply to inclusion of DAOs in the general category of
``securities" subject to regulation.\footnote{This reasoning would greatly
weaken the scope of the ``efforts of others" branch of \emph{Howey} as a
restrictive necessary condition for a token to be considered a security, and,
if pursued, most likely would trigger legal challenges and a possible
reconsideration or clarification of \emph{Howey} itself.}

It also is noteworthy that the SEC was deeply divided concerning the position
taken in the Reopening Release, with the minority group in the 3-2 Commission
split being sharply critical of that position. Commissioner Hester Peirce
articulated the critique in a dissent to the Reopening
Release.\footnote{\citet{Peirce_2023}.} She characterized the Release
position as an ``approach to exchange regulation as something that must not
-- indeed cannot -- be altered to allow room for new technologies or for new
ways of doing business."\footnote{Id.} In a later videotaped interview,
Commissioner Peirce took the position that the Commission should ``set true
decentralized protocols aside" because the ``technology itself is a
substitute for many of the regulatory protections" otherwise required when
interactions are with ``an intermediary, a centralized
firm."\footnote{\citet{Hamilton_2023}. \citet[pp. 2-3, 7-9]{DAOmodel2021}
describe this situation as one that involves ``functional and regulatory
equivalence.'' \label{regequiv} The application is the same as one that normally would be
regulated, functional equivalence, and elements in the technology achieve the
same purpose as regulation, regulatory equivalence.} She noted that you can
lose money and get hurt but ``you are opting into that when you are entering
into decentralized transactions."\footnote{Id.} This approach is very similar
in flavor to the newly enacted E.U. rule that exempts a crypto application
that operates ``in a fully decentralized manner without any intermediary."

A third regulatory approach, which can be termed \emph{regulatory
gradualism}, consists of observing a new technology and its operation and
then, after a period of observation, deciding on a new, appropriate
regulatory regime. Part of Commissioner Peirce's dissent to the Reopening
Release advocated this approach. She pointed out with approbation that the
SEC had used a regulatory gradualism approach in the 1990s, permitting new,
electronic exchanges to operate for a period of a few years despite being
non-compliant with the prevailing strict regulatory requirements for
exchanges, resulting in successful development and implementation of a new
body of regulations tailored to such exchanges. Closely related to this 1990s
example is the idea of creating a ``regulatory sandbox'' in which new
cryptocurrency technologies operate and within which participants understand
that they are on their own with respect to possible losses prior to the
imposition of an existing regulatory regime or the implementation of a new
one.\footnote{Commissioner Peirce has advocated this idea separately from her
dissent from the Reopening Release. See \citet{Piro2021}.} If the dissenters
gain a majority on the Commission in the future, policy may well move toward
this third approach or back to the first decentralization-focused approach
that would follow from classical adherence to \emph{Howey}.\footnote{The SEC
has five Commissioners appointed by the President, but only three can be from
the same party. The current 3-2 split is along party lines, with the
Republican Commissioners in the minority. Thus, the regulatory approach can
change sharply every time there is a Presidential election.}

\subsubsection{Interaction of the Mechanism with Regulation}

A key issue for both DAOs in general and for the mechanism in particular is the extent to which the associated technologies reduce the need for regulation. ``Regulatory equivalence" exists if elements in the technology achieve the same purpose as regulation.\footnote{See supra note \ref{regequiv}.} For example, to the extent DAOs and blockchain-based applications replace intermediaries with code, they eliminate the need for regulation to address potential misbehavior by such intermediaries.

Blockchain elements that typically accompany DAOs make not only the governing code public and transparent but also much if not all of the operations conducted by the DAO. This transparency reduces the need for many of the disclosures that regulators typically require in order to protect investors by giving them information about their investments. Requiring such disclosures is a major focus of regulation in the United States and many other jurisdictions.\footnote{The preamble of the Securities Act of 1933, one of the two key statutes that form a basis for securities regulation in the United States, describes the goals of the statute as follows: ``AN ACT To provide full and fair disclosure of the character of securities sold in interstate and foreign commerce and through the mails, and to prevent frauds in the sale thereof, and for other purposes." Securities Act of 1933, 15 U.S.C. §§ 77a-77aa (2018).}

In addition to requiring disclosure, regulatory systems typically include market structure and procedural protections for investors, often aimed at promoting fairness for such investors as they interact with sophisticated parties or exchanges.\footnote{A good statement of these purposes is the preamble of the Securities Exchange Act of 1934, the second in time of the two key acts aimed at protecting investors in the United States: ``AN ACT To provide for the regulation of securities exchanges and of over-the-counter markets operating in interstate and foreign commerce and through the mails, to prevent inequitable and unfair practices on such exchanges and markets, and for other purposes." Securities Exchange Act of 1934 (June 6, 1934, ch. 404, title I, § 1, 48 Stat. 881).} An example is the rules governing tender offers in the United States, offers that a bidder makes to shareholders of public companies to buy their shares for a particular price at a specified time. These rules require bidders to leave any offering open for at least 20 days, to refrain from giving different subscribers different terms, and to make certain disclosures.\footnote{15 U.S.C. § 78m; 17 CFR §240.13e, §240.14d, and §240.14e} Furthermore, management of the target corporation is required to issue its opinion of the offering terms.\footnote{Rule 14d-9 [17 CFR §240.14d-9] and Schedule 14D-9 [17 CFR §240.14d-101].}

The mechanism goes considerably beyond these customary disclosure-based, market structure, and procedural protections to provide direct substantive protection to investors. Consider, first, control period situations. Holding tokens at the beginning of an auction protects holders against any losses during the control period and guarantees that they will realize the increase in value or its financial equivalent promised by the control party on the portion of the tokens that they retain at the end of the auction. This arrangement exists whether or not the auction results in a shift in control as opposed to situations in which the winning bidder previously had control via an earlier auction or from a token position combined with active participation in governance during an open period. More generally, the auction aims to transfer as much of the social surplus from the control party's operations as possible to these token holders. This high degree of substantive protection in the face of shifts in control does not exist under current regulatory regimes. 

The control period protections are not operative for token holders during open periods and do not apply to tokens purchased during a control period after the auction has been completed.\footnote{The protection against losses and the guarantee of claimed gains applies to the particular holders at the time an auction is completed rather than attaching to the tokens themselves. Thus, buyers during a control period do not enjoy these protections. See supra page \pageref{signal}. Passive, ``buy and hold" investors enjoy particularly strong benefits overall because they always benefit from the control period protections.} But token holders in both situations enjoy two other protections. First, the threat of future auctions has a disciplinary effect with respect to misbehavior or poor performance on the part of existing governing parties, including the case in which a mass of diffuse token holders determine the DAO's path.\footnote{Because auctions may be freely initiated at any time and because the auctions are not subject to the restrictions that typically apply in corporate regulatory regimes, the threat implicit from auctions under the mechanism is most likely a much more effective external governance device than the threat from conventional corporate takeovers.} Second, these groups will enjoy future control period protections if subsequent auctions arise.

The substantive investor protections inherent in the mechanism both substantially strengthen the case for applying a less burdensome regulatory regime to DAOs that employ it and create the prospect of more fully protecting investors than existing regulatory regimes. They also make a DAO using the mechanism a very
attractive candidate for inclusion in a regulatory sandbox or for temporary
forbearance by regulators under regulatory gradualism approaches. 

There are additional ways in which the mechanism advances the case for a milder form of regulation as well as for regulatory gradualism. As we have seen, there is a strong argument that the mechanism enhances decentralization in the face of both implicit and explicit control possibilities, and in many cases actualities, that emerge from token ownership and voting regimes. The  mechanism induces choice of the path forward that will maximize social surplus and attempts to distribute that surplus in a manner that allows projects to reflect their true social value for purposes of initial and on-going investment. Excessive regulation would imperil these benefits. Finally, disclosure of the identity of the guiding parties of an enterprise is a major goal of securities regulation. The diffusion of control in DAOs tends to make identification of managing parties and elucidation of information about them difficult if not impossible. In contrast, by the operation of the mechanism, the identity of the managing parties will be evident, at least during control periods.

Do all of these features add up to a case for exemption? The answer to that question is not clear. There may remain some scope for regulation. In particular, disclosure of certain information beyond that available from the transparency inherent in blockchain aspects may have value to market participants and potential auction bidders. We leave exploration of that possibility and the question of exemption to future work.

Aside from creating a general case for reduced regulation or a regulatory gradualism approach, use of the mechanism raises four additional considerations with respect to existing regulation under the three approaches. First, there are the potential regulatory impacts from introducing the mechanism into DAO governance systems. Second, the mechanism creates regulatory opportunities, ways in which DAOs could more easily fit within existing regulatory regimes without adverse impacts on their mission and structure. Third, regulation may directly impact the effectiveness of the mechanism. Fourth, the mechanism raises regulatory questions if the DAO is subject to the core aspects of securities or commodities regulation.

The regulatory impact of shifting to the mechanism from the previous DAO
governance regime is unclear. For decentralization-focused approaches, the
impact of the mechanism on decentralization looms large, especially in Europe
where a finding that the DAO provides ``services ... in a fully decentralized
manner without any intermediary" would lead to exemption. As mentioned above,
the possibility of asserting temporary contestable control created by the
mechanism arguably enhances decentralization compared to current DAO
governance. At the same time, visible control during control periods creates
the opposite impression. The outcome is uncertain, especially given that the
E.U. rules are brand new.

In the United States, the creation of control parties by the mechanism can
lead to the claim that the ``efforts of others" branch of the \emph{Howey}
test is met because, in the words of \emph{Howey}, the control party's
efforts are the ``undeniably significant ones," essential to overall success
and profitability. Contestability likely is of no help here given the
disclosure rationale of the securities laws. Even if control is for a brief
period, investors would profit from knowing about the background and plans of
the control party in deciding whether to invest or divest. On the other hand,
it is clear that large token holders and parties that participate in implicit
control would be in the same boat under \emph{Howey}.

The mechanism creates some regulatory opportunities, ways in which a DAO may fit more easily within regulatory regimes without compromising its central purposes. At least during control periods, there is clear party who can coordinate and interact with regulators, parallel to the managers of a traditional corporation. That is not the case if there is only implicit control or a group of large token block holders. The ability of the control party to interact with regulators creates a shield for other parties such as developers, validators, and token holders. This shield may be especially valuable in the face of a joint and several responsibility regulatory approach such as the one suggested by the SEC in the Release that envisions a very broad set of actors being subject to potential regulatory obligations and liabilities. Otherwise, this approach might well, in the words of dissenting Commissioner Peirce, ``render innovation kaput."\footnote{\citet{Peirce_2023}.}

Some aspects of regulation may make it easier to implement the mech\-anism itself. As discussed above, one of the biggest challenges for the mech\-anism is toehold reporting. If the DAO is a security, then Schedule 13D reporting or a modified version of it would apply, creating a legal obligation to report a non-negligible toehold position.\footnote{\label{13(d)}Under section 13(d) of the Securities Act of 1934 any person who ``directly or indirectly" becomes the beneficial owner of more than 5\% of any class of registered equity securities, and certain equity securities that are exempt from registration must, within ten days, report certain information to the issuer, to any exchange on which the security is traded, and to the SEC. ``[I]f the purpose of the purchases or prospective purchases is to acquire control of the business of the issuer of the securities," the information must include ``any plans or proposals which such persons may have to liquidate such issuer, to sell its assets to or merge it with any other persons, or to make any other major change in its business or corporate structure." Securities Exchange Act of 1934, Pub. L. No. 73-291, 48 Stat. 881 (codified at 15 U.S.C. \S 78a (2012)). Schedule 13D is the form used to make the disclosures required by section 13(d).} This legal obligation and its enforcement is external, not a code feasible part of the mechanism itself. Nonetheless, it could go a long way toward filling what might be a significant implementation gap for the mechanism.

Even if the DAO is not considered a security under U.S. law, it will be
considered a commodity if it is publicly traded and thus subject to the
ability of the CFTC to police fraud. The DAO can leverage this potential
fraud liability with respect to the toehold position. Being able to submit an
artificially low bid by failing to report the entire toehold position is
fraud. The full enforcement and adjudication power of the CFTC comes into
play. The control party would be subject to discovery and required testimony
under oath. Hiding part of the toehold would create potentially serious
consequences for the control party. This potential fraud liability is another
external device, not part of the code feasible core of the DAO. But again, it
can help address a difficult implementation gap for the DAO.\footnote{It also
might be possible to create some structure in the DAO to aid the
applicability of this liability. For instance, the DAO might trigger notice
of an auction and the toehold reporting obligation to the CFTC, or, if the
DAO is a security, to the SEC, along with disclosure of the identity of the
control party. The CFTC and the SEC could incorporate this notice into their regulatory regimes as a trigger for
possible inquiry.}

The mechanism rules may have legal and regulatory implications if the DAO is
held to be a security or if it is an exchange subject to SEC or CFTC
registration and oversight. If the DAO is a security, then in the U.S.,
various state and federal laws covering corporate governance might apply.
Although the mechanism does not involve a tender offer or a merger, it is
possible that state corporate law requiring shareholder (here token holder)
votes for certain actions might apply to the auctions. The freeze-out feature
might be the subject of scrutiny even though it treats all of the $T_1$ token
holders the same. Jurisdictional issues are present. State law generally will
not apply unless the DAO is incorporated in the state. The ways in which this
class of rules apply and the extent to which they apply at all are complex,
and we leave full consideration to future work. The inquiry is important
because it also addresses the question of whether the mechanism here could be
used more generally rather than being limited to DAOs. In particular, the
strong points of the mechanism are equally cogent for traditional corporate
governance.

Finally, there are legal considerations in addition to regulation. Most
prominent is the issue of legal liability for DAO participants. This issue
exists with or without the mechanism. DAOs can be ``wrapped" in LLC or other
legal forms to provide limited liability for participants, including token
holders and any parties that might be seen as ``management."\footnote{See,
e.g., \citet{Guillaume_2022a}, \citet{Guillaume_2022b}, \citet{Kerr_2022},
\citet{Jennings_2022}, and \citet{Brummer_2022}.} Limited liability might be
particularly valuable for the control party due to the high visibility of
being in that position. That is the main new element created by the
mechanism. The question of the legal status and use of wrappers is a quickly
developing field and the choices are complex. We do not discuss that question
further here.

\section{Concluding Thoughts} \label{concluding}

We have shown that there is a plausible mechanism for DAO governance that is
code feasible, EV-robust, and efficient in the sense that it favors the
business plan with the highest social surplus. The auction aspect of the
mechanism provides a way out of the potential indeterminacy and pathologies
that arise from pure voting mechanisms. At the same time, the sequential
aspects of the mechanism create the ability to secure and enhance value that
stems from DAO voting and other non-market governance approaches. The fact
that empty voting is irrelevant to prevailing under the mechanism also
removes the possible arbitrariness or deliberate harm that might result. The
form of the mech\-anism permits the possible resolution of some major
regulatory dilemmas and creates a strong case for a reduced level of regulation that is nonetheless at least equally effective. The mechanism tends to minimize the added token value received by
project creators, thereby tending to maximize the investment value of the
project, albeit imperfectly. In the rest of this section, we consider the
mechanism in light of web3 and DAO governance norms.

One idealism of some web3 enthusiasts is that DAOs will create a new kind of
democratic community that is free from exclusive focus on token value or
domination by monied parties. It is not clear if the token value critique is
apt for economic DAOs, which are characterized by token value being a measure
of the value of the DAO activities. As discussed in the Introduction, cases
of economic DAOs extend way beyond DAOs that primarily have a commercial
purpose. Rather than being an evil, token value in the case of economic DAOs
creates the opportunity to provide superior governance through mechanisms
such as the one detailed here.

With respect to community, ``democratic" has many possible meanings. Here we
have taken the position that the goal is to maximize the value of each DAO
project with respect to an extended community that includes both present and
potential future participants. That value includes process value inhering
from ``democratic" aspects of DAO voting or other non-market DAO governance
mechanisms. If present or potential token holders value certain democratic
elements in excess of any associated losses from weaker operational
performance, the mechanism will favor those elements.

More generally, the mechanism permits an infusion of new participants and
approaches with respect to both DAO operation and governance. It is
impossible for large token holders, including founders or parties holding a
token majority, to retain control if there are parties who will outbid them
in an auction. That openness to innovators seems solidly in the spirit of
DAOs, especially given that the auction mechanism places them on an even
footing with the current governing interests in the DAO who also can bid and
win.

Observers have pointed out that many DAOs currently are ``undemocratic" due
to implicit control by founders or others who may have relatively large token
holdings while being surrounded by many small, disinterested holders of the
rest of the tokens. It may be that this situation is a tendency inherent to
governance through token voting, with implicit control shifting from one
group of such parties to another. In that case, DAOs are already
characterized by a succession of control parties, and the auction mechanism
only creates a more efficient and fair way to implement that succession.
Fairness here is not only with respect to potential control parties who are
freed from elements such as empty voting but also with respect to smaller,
more passive holders. These holders may be portfolio investors or may be less
active for Jensen-Meckling reasons: The benefits of costly effort to become
informed will accrue almost entirely to others. The mechanism encourages
value-increasing transitions and attempts to distribute a significant
proportion of the resulting social surplus to the passive holders.

It is important to note that the auction mechanism does not ban conventional
voting or vote buying regimes such as quadratic voting.\footnote{As discussed
in \citet{Lalley_2018}, when quadratic voting operates well, it aggregates
preferences and information accurately with respect to decisions. It is doubtful that quadratic voting could play a strong role in the context here in the face of asymmetric information and the possibility of free-riding. Parties without adequate information about potential control parties will not be able to represent their own interests well by voting, and revelation of information will trigger free-riding. If a project creator expends a large amount in a quadratic vote, the project creator will not be able to cover it from added token value due to that free-riding. If the DAO returns amounts the creator paid for the votes required to prevail to the creator, voting
incentives are distorted.\\
\indent The mechanism described here fits solidly within a category that \citet[p. 36]{Lalley_2018} describe as ``the natural private goods markets analog of the radical market logic of [quadratic voting]," namely ``assets ... being continuously auctioned for rental according to ... [an] English auction." Under the mechanism, there is no continuing private property right in controlling the DAO. Instead, it is sequentially ``rented out" to winning auction bidders, those who offer to pay the highest amount of ``rent" in the form of part or, ideally, all of the social surplus they can generate during a temporary ``tenancy" period.  \citet{Lalley_2018} associate this kind of approach with Henry George and William Vickery. } These approaches and
the ensuing wisdom of crowds or attractive aggregative properties would
operate fully during the open periods interspersed among control periods.
Furthermore, even during control periods, the control party can abstain
partially or fully on particular votes in order to allow the aggregative
benefits of voting to be realized based on the control party's calculus of
how much expertise it has on those votes compared to the mass of other
voters.\footnote{See \citet{Bar_Isaac_2020}.} Some DAO projects already
conform to a rigid version of this more flexible possible control period
behavior. A foundation or other centralized element implicitly or explicitly
makes certain decisions with respect to the DAO while others are left
entirely up the decentralized DAO governance mechanism. The auction mechanism
here could extend to some current projects as a unified whole, encompassing
both the centralized and decentralized portions of the projects. This
approach would create on-going flexibility to allocate decisions between 
voting mechanisms and more centralized methods in the overall enterprise by using control periods to restructure the balance between the on-going centralized and decentralized elements. This ability to renew and redesign the structure in a coherent way would be perpetually available.

That perpetual ability to renew and redesign the unified enterprise also
addresses another rigidity, one that is intertemporal. Many enterprises begin
with centralized incubation of the DAO element and then reach a point where
governance passes entirely to a ``fully decentralized'' state, relying only
on the DAO governance mechanism.\footnote{MakerDAO is a prominent example.
Maker Foundation was a centralized entity used to develop the MakerDAO
system. As described in \citet{MakerWP_2020}, the Maker Protocol Whitepaper:
\begin{quote} The Maker Foundation currently plays a role, along
with independent actors, in maintaining the Maker Protocol and expanding its
usage worldwide, while facilitating Governance. However, the Maker Foundation
plans to dissolve once MakerDAO can manage Governance completely on its own.
 \end{quote}
In July 2021, MakerDAO reached a point described by the Maker Foundation CEO
as being ``completely decentralized," with the transfer of all Foundation
functions to the DAO and the subsequent envisioned formal liquidation of the
Foundation. \citet{Christensen_2021}. \label{Maker}} One potential serious problem with
this approach is that the enterprise may benefit at a later point from a
major reorganization or revamp, perhaps with innovation in mind, that is
critical to maintaining on-going dynamism or even for survival of the
project.\footnote{Exactly this situation appears to have emerged for MakerDAO. Only a short time after it presumably became completely decentralized, see note \ref{Maker} supra, Rune Christensen, one of the co-founders, used his large token position to initiate a major reorganization of MakerDAO through a project called ``Endgame." The goal was to bring MakerDAO to the point where it could operate at a mass market level. Because of Christensen's dominant token position, implementation of Endgame came about thorough an effective re-centralization of the project, a step that Christensen envisioned as temporary. See \citet{GilbertAk_2023}.}  This re-organization or revamp may be difficult or impossible to
accomplish without reinstating a later period of centralization that
incubates a rebirth of the DAO in a renewed and more valuable
form.\footnote{Relevant in this regard is Vitalik Buterin's lament concerning
Ethereum in 2022. \citet{Buterin_2022}. Buterin stated that he had
``diminishing influence" over Ethereum and that it was ``becoming harder to
make big changes to the Ethereum protocol due to the many stakeholders that
have a say in the decision-making process," creating an Ethereum that was
``definitely more vetocratic." He stated that: ``Even now, I feel like the
window is closing on substantial things. It's getting harder to do big things
even today." Ethereum is not governed by a DAO, but the possibility of
getting to a stale point is equally applicable to DAOs. Reorganization and major
revamps are frequent in the corporate and  non-profit world. It is not clear
why DAOs or Ethereum would be any different with respect to the possible
benefits of occasional major overhauls.} It is very hard to see why
restricting centralization to the beginning of the project is optimal in
general. The auction mechanism leaves open the possibility of major reform
during a control period by enabling consistent and coherent execution of a
new plan similar to the plans used to birth DAOs during an initial period of
centralized incubation. A control period for major reform can be invoked
through the mechanism on a when-needed basis, with the most promising plan
the odds-on favorite to be implemented.

We believe that the mechanism described here fits well within the ideals that have
inspired DAOs. Control over the direction of the DAO under the mechanism is
continuously contestable. Founders and other parties no longer can maintain
an iron grip through implicit control. The DAO is safe for democracy of
various kinds because an auction followed by a control period can correct the
adverse consequences of voting mechanisms if necessary, defeat empty voting
if it becomes a problem, and overturn attempts of malicious actors to damage
the DAO. The interests of passive small holders are protected. The value of
the DAO project is maximized to the benefit of all participants. Constructive
innovation is facilitated.

The goal here has been to present one possible mechanism that has promise. No
claim is made that this mechanism is the best one. We have suggested
alternatives for a variety of mechanism aspects throughout. If this kind of
auction-based approach is found to be promising, there is much more work to
be done.

\appendix

\section{Appendices}

\subsection{Proofs}

We construct a proof of Proposition 1 and Corollary 2, proving Lemma 1 and Corollary 1 near the end of the main proof when the Lemma is needed.

Suppose that a potential bidder can implement a business plan that involves
expending effort, consisting of labor and resources equivalent to $C$
monetary units, in order to increase the value of the DAO by $N(C)$ monetary
units. Define $V = V(C) = N(C)/q +P_0$ to be the token value emerging from
this business plan. The basic auction is a vehicle for the potential bidder
to gain control in order to implement the plan. We compute the strongest
potential bid, $(S, R, t_b)$, the one that maximizes $A=(1-t_b)(S-P_0) q -
R$, based on the plan.

The total social surplus generated by the plan is $\psi = (V-P_0) q -C$. Consider first the case $S \le V$. In that case, the
bidder's profit function conditional on winning the auction based on a bid
$(S, R, t_b)$ is:
$$\Pi_b = t_b(V -P_0) q -C + \frac{R}{S-P_0}(V-P_0).$$
The first term, $\Pi_{t_b} = t_b(V-P_0) q$, is the bidder's expected profit on
the bidder's toehold tokens, which have value $P_0$ at the time of the bid.
The second term is the bidder's cost. The third term, $\frac{R}{S-P_0} (V-P_0)
= t_f q (V-P_0)$, is the bidder's expected profit on the proportion $t_f$ of
the total $q$ tokens for which the bidder has forced purchase or voluntarily
sold at price $P_0$ using the freeze-out feature of the auction mechanism.

To create the best possible bid,
the bidder will choose $R = \frac{S-P_0}{V-P_0}\left(C-\Pi_{t_b}\right)$
which is the smallest $R$ subject to the constraint of breaking even,
$\Pi_b=0$. Assume for the moment that this value of $R$ satisfies the market
size constraint $t_f \leq t_m - t_b$ required for a valid bid. Then:
\begin{equation*}
\begin{split}
A & = (1-t_b)(S-P_0) q - R\\
& = \frac{S-P_0}{V-P_0} \left[ (1-t_b)(V-P_0)q - C +
\Pi_{t_b} \right]\\
& = \frac{S-P_0}{V-P_0} \left[ (V-P_0) q - C \right]
\end{split}
\end{equation*}
Clearly $A$ is maximized by the choice $S = V$, which yields:
\begin{equation}  \label{eq:opt_bid_VC}
A =  \left[ (V-P_0) q - C \right].
\end{equation}
$A$ is the total social surplus from the business plan, which equals the social gain
from implementing it. It represents the strongest bid \emph{conditional} on
adopting this business plan.

But there may be a better business plan that produces even higher social surplus,
and $A^*$, the best overall bid, will emerge from maximizing total social surplus,
$\psi(C) = V(C) q - C$. Assuming that $V(C)$ is twice continuously
differentiable, that $V^\prime(0)> 0$, and that $V \dblprime(C) < 0$, there
will be a maximum defined by $C^* = \arg \max_{C} V(C) q - C$, and $V^* =
V(C^*)$. The strongest possible bid from a bidder characterized by the value
function $V(C)$ is:
\begin{equation} \label{eq.opt_bid_V*C*}
A^* =  \left[ (V^*-P_0) q - C^* \right].
\end{equation}

So far we have ignored the possibility that a pair $(V,C)$ and, in particular,
$(V^*,C^*)$ is infeasible because it violates the market size constraint
$t_f \leq t_m - t_b$ required for a valid bid. In the case of a break-even
bid, violating this constraint means that the bidder cannot acquire enough
tokens through the freeze-out feature combined with the bidder's toehold
position, if any, to cover the cost $C$ through added token value realized on the
bidder's token position. The bidder can acquire at most the proportion $t_m$
of the total tokens by these means. Note that if $t_m = 1$, violation of the
constraint implies that $\psi(C) < 0$. In that case, the business plan
destroys value.

For a pair $(V,C)$, the highest possible break-even bid requires $R = C - (V
- P_0) t_b q$, i.e., proposing an $R$ that just covers the cost $C$. This
choice of $R$ along with $S = V$, which is necessary for the break-even bid
to be the highest, result in:
\begin{equation*}
\begin{split}
t_f  & = \frac{R}{(V-P_0) q}\\
& = \frac{C}{(V-P_0) q} - t_b.
\end{split}
\end{equation*}
The market size constraint becomes:
$$(V(C)-P_0)q \geq \frac{C}{t_m}.$$
If the constraint is binding, then $V^\prime(C) = \frac{1}{t_m q}$ versus
$V^\prime(C) = \frac{1}{q}$ in the case of unconstrained optimization. The
higher value of $V^\prime(C)$ implies lower values for $V^*$, $C^*$, and the
social surplus, $\psi^*$, in the constrained optimization. Equation
\eqref{eq.opt_bid_V*C*} still applies, but with these lower values. Given a
lower value for the social surplus $\psi^*$, the consequent bid, $A^*$, also will be
lower.

Suppose that the bidder in fact can make the highest break-even bid $A_1^*$
of any of the auction participants and that the second highest break-even bid
is $A_2^*$. Then the bidder can capture $\psi_e =\min \{(1-t_m)\psi^*,
A_1^*-A_2^* \} $ of additional social surplus from the business plan $(V^*,C^*)$.
The bidder accomplishes that by adding $\psi_e$ to $R$ in the bid as if it
were an additional cost, which will reduce the bid by $\psi_e$.\footnote{The
bidder can increase the resulting reduced bid by a small amount $\epsilon$ if
there is a need to defeat a competing bid of exactly $A_2^*$.} Retaining the
business plan $(V^*,C^*)$ and adjusting $A_1^*$ as indicated dominates
changing the business plan and submitting a different bid based on the
alternative plan. Because the alternative plan will result in total social surplus
$\psi_a < \psi^*$, a bid based on the alternative plan that directs the same
amount of social surplus to the bidder will be lower than the corresponding bid
under the original plan.

Now consider the case $S > V$.  We will contrast this case with the previous
one in which the optimal value claim was $\bar{S} = V$ accompanied by a surplus
claim of $\bar{R}$ and an auction parameter $\bar{A}$. The value claim $S>V$ will cause the bidder to lose $\phi(D_v,P_0,S,V)$ from the functioning of the value deposit forfeit function because only the level $V < S$ is attainable
through the bidder's efforts. As a result, the bidder's profit function will include an additional negative term compared to the $S \le V$ case, becoming:
$$\Pi_b = (V -P_0)t_b q -C + \frac{R}{S-P_0}(V-P_0)-\phi(D_v,P_0,S,V)$$
Furthermore, to maintain the same amount of
added token value that would accrue with a value claim of $\bar{S} = V$, the bidder will
need to secure the same value of $t_f$ by scaling $\bar{R}$, the surplus claim
accompanying the value claim $\bar{S} = V$, by the factor
$\frac{S-P_0}{V-P_0}$. The bidder also will need to increase the surplus
claim by $\frac{S-P_0}{V-P_0}\phi(D_v,P_0,S,V)$ in order to cover the
anticipated value deposit loss. The condition in Lemma 1 requires that $\phi(D_v,P_0,S,V) > (1-t_d)(S-V)q$. Define $\delta(S,V) = \phi(D_v,P_0,S,V) - (1-t_d)(S-V)q$.

We compute the surplus claim $R$ required in the case $S>V$ for the bidder to realize the amount of added token value $\bar{R}$ that was claimed in the case $S = V$ assuming in both instances that the business plan results in token value $V$:
\begin{equation*}
\begin{split}
R & = \frac{S-P_0}{V-P_0}\left[\bar{R} + (1-t_b-t_f)(S-V)q + \delta(S,V)\right]\\
& = \frac{S-P_0}{V-P_0}\left[\bar{R} + (1-t_b)(S-V)q + \delta(S,V)\right]-\frac{S-V}{V-P_0}R\\
& = \bar{R} + (1-t_b)(S-V)q + \delta(S,V).
\end{split}
\end{equation*}
\noindent where the first step uses $t_f=\frac{R}{(S-P_0)q}$ and the second step uses $1 + \frac{S-V}{V-P_0} =  \frac{S-P_0}{V-P_0}$.\\

With the required R in hand, the bid $A$ that will produce the same amount of expected bidder added token value as the bid $\bar{A} = (1-t_b)(V-P_0)q-\bar{R}$ is:
$$A = (1-t_b)(S-P_0)q-\left(\bar{R} + (1-t_b)(S-V)q + \delta(S,V)\right)$$
\noindent and the change in the bid holding expected bidder added token value constant is:
\begin{equation*}
\begin{split}
\Delta A & = A - \bar{A}\\
& = -\delta(S,V).
\end{split}
\end{equation*}
$V$ is private information of the bidder, not known by the mechanism designer. Including a value claim with $S > V$ will be a dominated strategy for any value of $V \in [P_0,S]$ if and only if $\delta(S,V) > 0$ for all values of $V$, establishing Lemma 1. Corollary 1 follows because $D_v \ge \phi(D_v,P_0,S,P_0) = (1-t_d)(S-P_0)q + \delta(S,P_0)$ and $\delta(S,P_0) > 0$.

\subsection{Hedging and Stochastic Adjustments}\label{appendix:stochastic}

We developed the sequential auction mechanism in a deterministic setting
where changes in the token price are driven solely by winning bidder business
plan innovations. We now assess the mechanism in a stochastic setting, first
discussing winning bidder hedging strategies and then possible mechanism
adjustments that, in effect, provide hedging at the DAO level.

Honest bidders, those that are not aiming to engage in strategies such as
value destruction, enter an auction because they believe they can enhance the
value of the DAO and potentially profit from those enhancements themselves.
Doing so in a stochastic environment involves taking a risky token position
and exposing themselves to potentially losing part or all of their value
deposit or surety deposit due to token price fluctuations not associated with
the effectiveness of their efforts to add value. In addition, unless
offsetting adjustments are made, these \emph{stochastic elements} will tend
to create imbalances in winning bidders' portfolios, especially if the
required position in the DAO is large relative to the other positions in
those portfolios. As a result, winning bidders will want to hedge the
position they are taking in the DAO. We sketch how the hedging might operate,
leaving aside a more formal development in favor of creating an accurate but
more intuitive perspective.

Suppose that absent the winning bidder's project, the token price would
evolve as a random variable, $\widetilde{W}_b$, the \emph{base outcome}, and
that the outcome of the winning bidder's efforts changes the random variable
to $\widetilde{W}_b + \widetilde{W}_e$ where $\widetilde{W}_e$ is the
\emph{effort outcome}, with $E(\widetilde{W}_e) = V^* - P_0$. Typically
$\widetilde{W}_e$ will not be stochastically independent of
$\widetilde{W}_b$. For example, if a high outcome of $\widetilde{W}_b$
occurs, it also might mean that the winning bidder's effort is more
productive in creating value, with a consequent high outcome for
$\widetilde{W}_e$. The past history of DAO token prices creates insight into
$\widetilde{W}_b$. But as of the end of the auction, there is no history for
$\widetilde{W}_e$, and its qualities, to the extent surmisable, are private
information of the bidder.

These two random variables each contain both systemic and idiosyncratic
elements. The market prices systematic risk components, ones that cannot be
diversified away, but does not price idiosyncratic risk, which can be
diversified away because it nets out to zero across the market portfolio.

The identity of the winning bidder matters. Consider two extreme cases. On
the one hand, there is a ``bare bones" entrepreneur with limited resources
and wealth, who will be unable to diversify away idiosyncratic risk and who
may have difficulty fully hedging systematic risk. The danger is that this
entrepreneur will forgo undertaking a project with high social surplus
because of the accompanying high risk and portfolio distortions.

On the other hand are institutional investors, such as a hedge funds, private
equity firms, or venture capital firms that exist in part because they are
vehicles to solve exactly these hedging and diversification problems. These
firms typically create portfolios of investments and themselves are owned by
other investors, sometimes large pools of them. Idiosyncratic risks are
thereby shifted to the broader investment system and can be diversified away.
Systematic risk is absorbed by outside parties in the broader investment
system hedging or adjusting their portfolios. We begin the analysis with the
winning bidder being an institutional investor and then consider the case of
a bare bones entrepreneur.

There are three separate elements of concern with respect to hedging and
diversification for a winning bidder: (i) the token position consisting of
$t_d q$ tokens; (ii) the value deposit liability; and (iii) and the surety
deposit liability. An institutional investor will simply absorb the token
position. The idiosyncratic risk will be diversified away and the systematic
risk shifted to appropriate parties through the broader investment system.
The institutional investor can focus purely on the question of choosing a business plan that maximizes social surplus creation, aligning their motivation exactly with the auction mechanism. The deterministic analysis applies directly, interpreting $V^*$ as an expected value rather than a certainty.

The same expected value relationship does not apply for the value deposit and
surety deposit. These deposits function as derivatives positions under the
associated DAO Code rules. The value deposit is analogous to a bearish put spread, the analogy being exact in the case of the standard token deposit forfeit function with no baseline loss penalty.\footnote{See supra note \ref{options note} and the accompanying text.} With respect to outcomes below $P_0$, a large put position with appropriate strike prices would be required to offset potential forfeits involving both the value deposit and surety deposit. The $T_1$ token holders, who benefit if the deposits are forfeited in part or entirely by the control party, effectively hold all these derivatives positions. When the control party succeeds in bringing the token value up to $V^*$ in the deterministic setting, both deposits are fully refunded and the $T_1$ token holders receive nothing. In a stochastic setting, $V^*$ is an expected value, not a certainty. There is a probability that the token price will never reach $V^*$ due to stochastic elements unrelated to the cogency of control party's business plan and the effectiveness of execution of the plan. In that case, the control party will lose part or all of the value deposit, and if the stochastic elements are strong enough on the downside to result in a token price below $P_0$, part or all of the surety deposit. There is no corresponding upside: The deposits are fully refunded if $V^*$ is reached on a sustained basis, but attaining a higher level than $V^*$ results in no additional benefit. The overall result is that in the stochastic case, the implicit derivatives positions inherent in the two deposits both have a negative expected value for the winning bidder, while in the deterministic case they had zero expected value.

As in the case of the token position, an institutional investor can pass on
the idiosyncratic risk and systematic risk associated with the implicit
option positions arising from the deposit arrangements to the broader
investment system, which can eliminate the idiosyncratic risk and absorb the
systematic risk. But the institutional investor cannot avoid the negative
expected value hit associated with these positions, a hit that artificially
lowers the appeal of surplus-creating business plans. Adjustment of the
auction mechanism itself is required to address this problem.

Adjustment, at least to a good approximation, is possible in a code feasible
way. The stochastic characteristics of the token price, $\widetilde{W}_b +
\widetilde{W}_e$, such as volatility are key to the adjustment. With elements
such as variance and skewness in hand, one can calculate the expected value
of the loss on both implicit derivatives positions and reduce the two
deposits accordingly along with appropriate changes in the rules for
returning the deposits. However, as discussed and summarized in
\citet{Khotari2007}, these elements must be estimated, are not stable over
time, and post-event values are superior for purposes of estimation. The
superiority of using post-event rather than pre-event values is particularly
clear in the case of the basic auction because the token price process shifts
from $\widetilde{W}_b$ to $\widetilde{W}_b + \widetilde{W}_e$, a shift that
typically will alter the distribution of returns, including the volatility
parameters relevant to making adjustments. Because the control party has an
interest in the outcome of the estimation, an estimation method would need to
be hard-coded at the constitutional level of the DAO to avoid manipulation by
the control party. Oracles might be necessary to retrieve the necessary data.
Given the data and protection from manipulation, estimating basic volatility
parameters seems quite feasible, at least to a good approximation.

These estimation difficulties are only a small taste of the problems that
would attend deeper mechanism revisions to address stochastic elements. At
the highest level of ambition, the mechanism would isolate the impact of the
new business plan by estimating the value impact of that business plan. A
classic way to perform this kind of estimation is an event study.
\citet{Khotari2007} point out that event studies can be fairly accurate and
robust in the short run but that they are unreliable in the long run, often
failing to detect large effects at all in simulations, much less determine
their magnitude accurately. Thus, even if embodying event study methodology
in a DAO were code feasible, it is not likely to be effective. It certainly
would fall short of letting the bulk of the adjustments occur externally through institutional control parties shifting risk to the general market, leaving only the relatively easy adjustments for the intrinsic option positions to the DAO Code.

We now turn to the opposite polar case in which the potential bidder is a
bare-bones entrepreneur instead of an institution that provides flow through
to the general investment market and has deep resources. We have highlighted
the danger that the bare-bones entrepreneur may choose to forgo a project
with high social surplus due to unacceptably high risk. The entrepreneur also
may need substantial funding to cover the value and surety deposits as well
as the cost of the business plan. A classic market solution to these problems
is to put the entrepreneur in a management role, with one or more intuitional
investors providing the funds, and then to create a management contract that
gives the entrepreneur incentives to choose and implement the highest
value-added project net of cost, where the cost itself is covered by investor
funding. The investors are diversified and are not affected by idiosyncratic
risk. In this context, \citet{Park2015} discuss the design of management
incentives through compensation contracts, summarize much of the previous
literature, and show that a management contract that includes convex
compensation such as call options is superior if it is set up with a
diversified index as the baseline. Call options reward the manager for
achieving the high token values envisioned in the business plan and, combined
with salary and other stable elements, align the manager's financial returns
with attaining success with the business plan. The diversified index baseline
moves the focus of risk-taking to idiosyncratic risk, which is immune to
manager hedging. Suitable levels of the call option compensation will create
the required incentives for the manager to fully execute on the business plan
despite the attendant risks.

\newpage
\bibliographystyle{econ_no_doi} 
\bibliography{Econ_and_Game_Theory}




\end{document}